\documentclass[
    aps,
    prl,
    twocolumn,
    floatfix,
    superscriptaddress
]{revtex4-1}

\usepackage[utf8]{inputenc}
\usepackage{times}

\usepackage[breaklinks, unicode]{hyperref}
\usepackage{url}

\usepackage{siunitx}
\usepackage[version=4]{mhchem}

\usepackage{amsfonts}
\usepackage{amssymb}
\usepackage{amsthm}
\usepackage{mathtools}
\usepackage{braket}

\usepackage{tikz-cd}
\usepackage{xcolor}
\usepackage{graphicx}
\usepackage{float}

\newcommand{\pgi}{Peter Gr\"unberg Institut and Institute for Advanced Simulation,
Forschungszentrum J\"ulich and JARA, 52425 J\"ulich, Germany}

\newcommand{\aachen}{Department of Physics, RWTH Aachen University, 52056 Aachen, Germany}

\newcommand{\mainz}{Institute of Physics, Johannes Gutenberg University Mainz, 55099 Mainz, Germany}

\newcommand{\NCG}{noncommutative geometry}

\newcommand{\Algebra}{\mathbb{A}_\hbar}

\newcommand{\Conn}{\mathcal{A}}
\newcommand{\Curv}{\mathcal{F}}
\newcommand{\conn}{A}
\newcommand{\curv}{F}

\newcommand{\Qmetric}{\mathrm{g}}

\newcommand{\ad}{\mathrm{ad}}

\newcommand{\cov}{\nabla}

\renewcommand{\vec}[1]{\mathbf{#1}} 

\newcommand{\hatn}{\hat{\mathbf{n}}}

\newcommand{\abs}{\mathrm{\abs}}
\newcommand{\wigner}{\mathcal{W}}

\newcommand{\lpartial}{\overset{\leftarrow}{\partial}}
\newcommand{\rpartial}{\overset{\rightarrow}{\partial}}

\newcommand{\tr}{\mathrm{tr}}
\newcommand{\Tr}{\mathrm{Tr}}

\newcommand{\fraktr}{\mathfrak{Tr}}

\newcommand{\diag}{\mathrm{diag}}

\newcommand{\sgn}{\mathrm{sgn}}

\newcommand{\id}{\mathrm{id}}

\newcommand{\dd}{\mathrm{d}}

\newcommand{\volume}{\mathcal{V}}

\newcommand{\Gk}{\underline{G}}

\newcommand{\Gret}{G^{\mathrm{R}}}

\newcommand{\Gadv}{G^{\mathrm{A}}}

\newcommand{\bsigma}{\boldsymbol{\sigma}}

\newcommand{\xc}{\Delta_\mathrm{xc}}

\newcommand{\soi}{\alpha_{\mathrm{R}}}

\newcommand{\spacetime}{{\mathbb{R}^{1,d}}}

\newcommand{\figureI}{
\begin{figure}[t!]
    \centering
    \includegraphics[width=\linewidth]{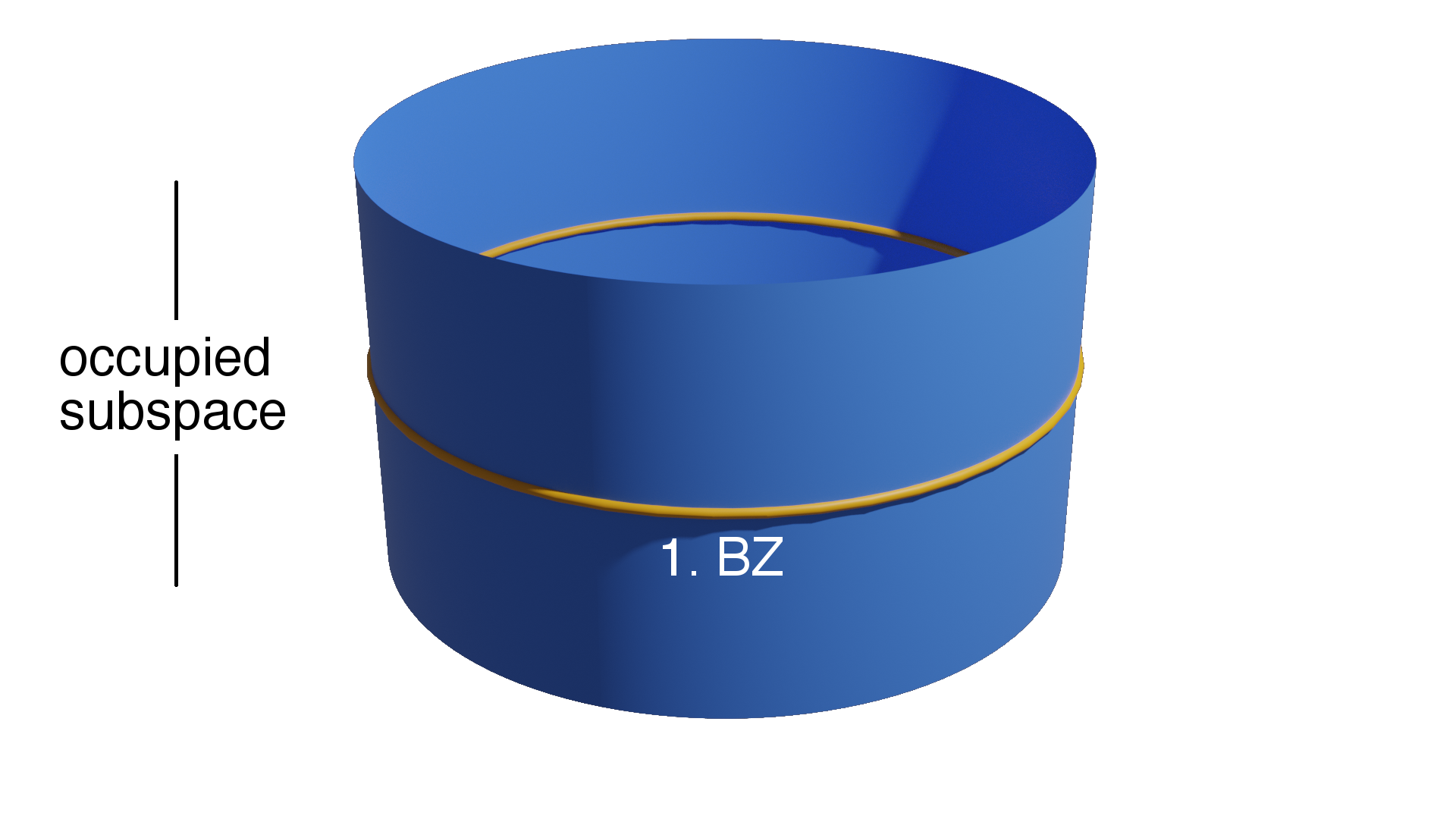}
    \caption{Illustration of the Bloch bundle. By diagonalizing the Hamiltonian of an insulator, one can assign a projection operator $\varrho_\vec{k}$ to each $k$-point in the first Brillouin zone (1.BZ; here illustrated as a circle in one dimension). One obtains a projective module: states at $\vec{k}$ which are invariant under the action of $\varrho_\vec{k}$, belong to the occupied subspace (here illustrated as the cylindrical surface for a one dimensional occupied subspace). In this sense, a projective module gives rise to a vector bundle in classical geometry. }
    \label{fig:bloch_bundle}
\end{figure}
}

\newcommand{\figureII}{
\begin{figure*}
    \centering
    \includegraphics[width=1\linewidth]{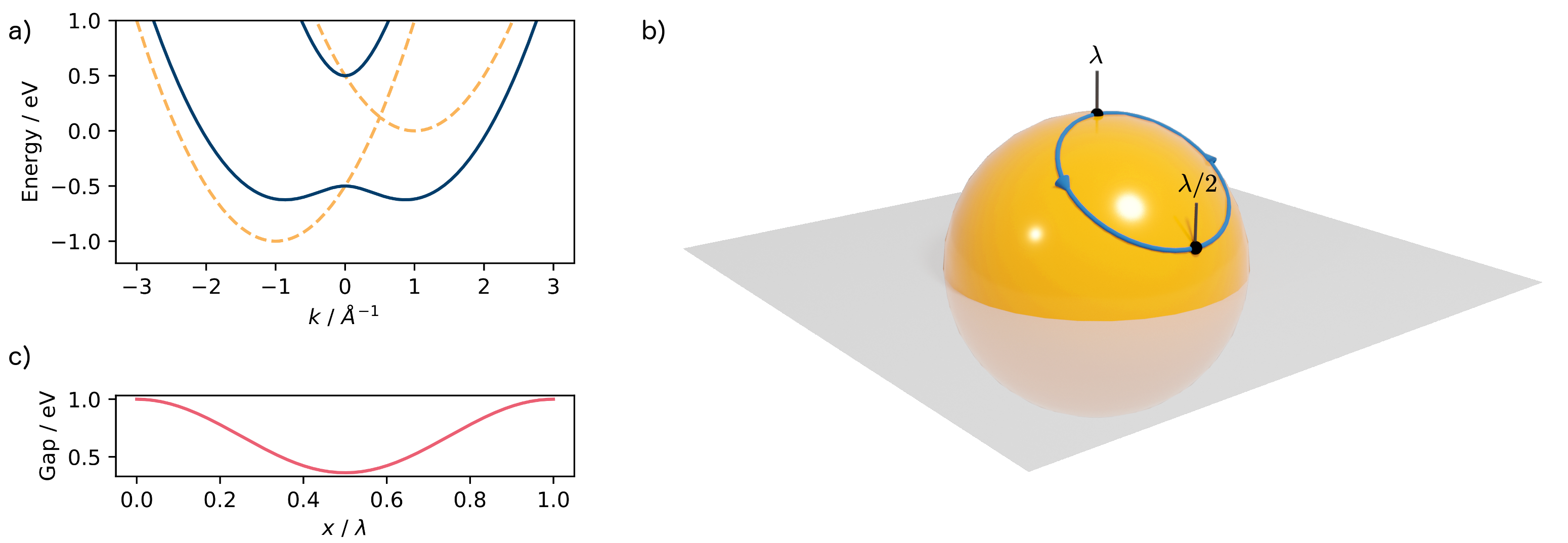}
    \caption{Band structure evolution in the magnetic Rashba model. Fig. a) displays the band structure in the two-dimensional Rashba model for two different magnetization directions:
    $\hatn=\vec{e}_z$ (solid line) and a configuration with $n_z = 0$ (dashed line). The bands are nondegenerate for all momenta only when $n_z \neq 0$. Whenever $n_z=0$, the bands cross at a singular point $\hbar\vec{k}_c = (-n_y, n_x)^T \xc / \soi$  which corresponds to a topological phase transition. The plot shows a line cut of the band structure along this critical direction. Imagining a conical spin spiral phase of wavelength $\lambda$, the local magnetization directions would trace a circular path when mapped to the unit sphere which is shown in Fig. b). By carefully choosing the parameters of the spin spiral, the system is driven close to its topological phase transition (gray plane) while never quite reaching it. This can be quantified in Fig. c), by following the evolution of the band gap at $\vec{k}_c$ along the direction of the spin spiral. The bands are driven close to degeneracy at $x=\lambda/2$, where strong Berry curvature effects can be expected.}
    \label{fig:bandstructure}
\end{figure*}
}

\newcommand{\figureIII}{
\begin{figure}[t]
    \centering
    \includegraphics[width=\linewidth]{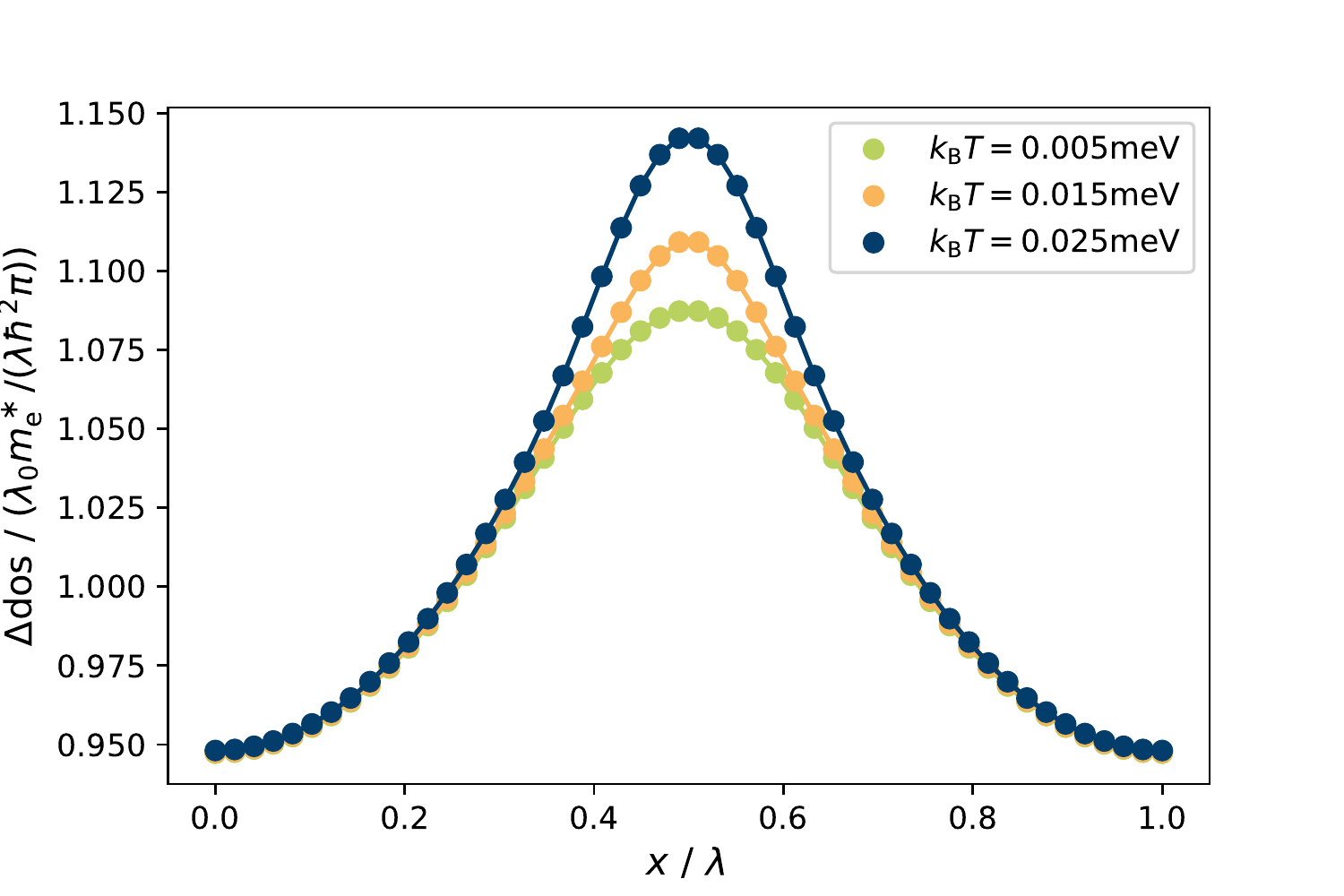}
    \caption{Variation of the local density of states (DOS) for a conical spin spiral. The DOS is normalized to its value in a free electron gas: $m^\ast_\mathrm{e} / (\pi \hbar^2)$. The wavelength of the spiral is given by $\lambda$, while a reference length is set to $\lambda_0= \SI{1.95}{\angstrom}$. The model parameters are: $m^\ast_\mathrm{e} = m_\mathrm{e}$, $ \xc = \SI{0.5}{\electronvolt}$, $\hbar\soi= \SI{1.95}{\electronvolt \angstrom}$ }
    \label{fig:dos}
\end{figure}
}

\newcommand{\figureIV}{
\begin{figure}
    \centering
    \includegraphics[width=\linewidth]{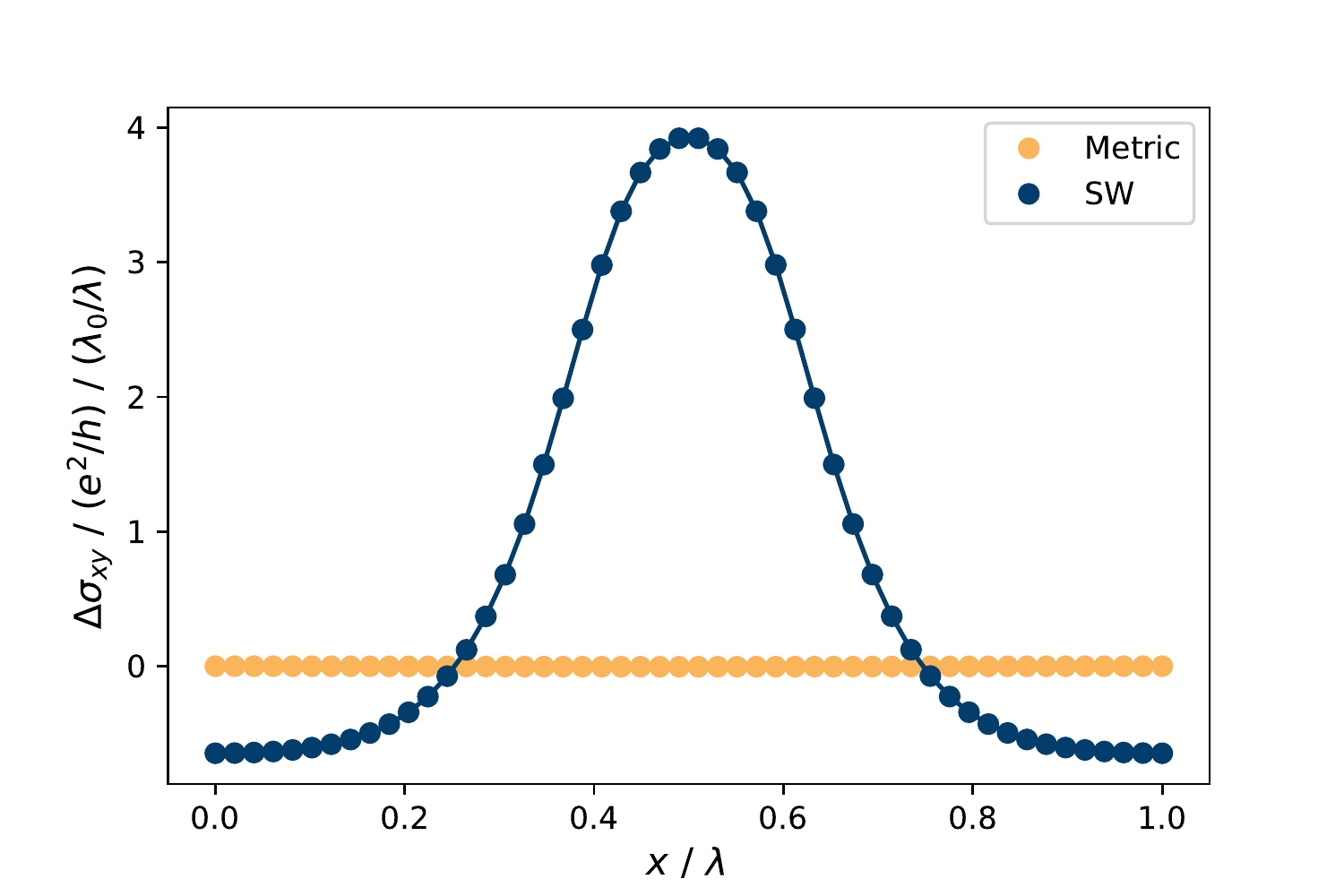}
    \caption{Local variation of the anomalous Hall conductivity for a conical spin spiral. The wavelength of the spiral is given by $\lambda$, while a reference length is set to $\lambda_0= \SI{1.95}{\angstrom}$. The model parameters are: $m^\ast_\mathrm{e} = m_\mathrm{e}$, $ \xc = \SI{0.5}{\electronvolt}$, $\hbar\soi= \SI{1.95}{\electronvolt \angstrom}$ and $k_\mathrm{B} T = \SI{25}{\milli\electronvolt}$.}
    \label{fig:chiral_hall_effect}
\end{figure}
}

\begin{document}
%====================================================

\setcounter{secnumdepth}{2} 
\hbadness=2000 

%-- Header  -----------------------------------------

\title{
    Effective Seiberg-Witten gauge theory of noncollinear magnetism 
}

\author{Fabian R. Lux}

    \email{f.lux@fz-juelich.de}
    \affiliation{\pgi}
    \affiliation{\aachen}
    
\author{Pascal Prass}

    \affiliation{\mainz}
    
\author{Stefan Bl\"ugel}

    \affiliation{\pgi}
    
\author{Yuriy Mokrousov}

    \affiliation{\pgi}
    \affiliation{\mainz}

\date{\today}

\begin{abstract}
Smoothly varying magnetization textures such as domain walls, skyrmions or hopfions serve as promising candidates for the information bits of the future. Understanding their physical properties is both a major field of interest and a theoretical challenge, involving the physics on different length scales.
Here, we apply the phase space formulation of quantum mechanics to magnetic insulators and metals in the limit of zero temperature to obtain a gradient expansion in terms of real-space derivatives of the magnetization.
Our primary focus is the anomalous Hall effect in noncollinear magnets which serves as an important proxy in the detection of localized magnetic structures.
We formulate the problem in the language of noncommutative fiber bundles and make the central finding that the semiclassical expansion of the density matrix and the Berry curvature is governed by a construction from string theory which is known as the Seiberg-Witten map.
Originally discovered in the effective low-energy behavior of D-branes, this map now gives a geometrical underpinning to gradient expansion techniques in noncollinear magnets and offers a radically new perspective on their electronic properties.
\end{abstract}

\maketitle

%----------------------------------------------------
\section{Introduction}
%----------------------------------------------------

The fundamental principles of geometry and topology permeate all areas of modern physics~\cite{Nakahara2003, Frankel2011} and also the field of condensed matter offers its own prime manifestation: the integer quantum Hall effect (IQHE)~\cite{VonKlitzing1980, VonKlitzing1986, Thouless1982, Kohmoto1985}.
When subjected to strong magnetic fields, a two-dimensional electron gas will develop topologically protected states whose hallmark is a quantized transversal conductivity, robust even in strongly disordered systems.
Soon after its discovery, it was realized that this topological effect has its roots in a flavor of modern geometry known as  \emph{noncommutative geometry}~\cite{Bellissard1994}. 
A novel field at the time, noncommutative geometry has occupied an important role in various fields of physics ever since its conception, notably in high energy physics and string theory~\cite{Seiberg1999, Connes1999, Douglas2001,Szabo2003}.

For ordinary, everyday, compact spaces the so-called Gelfand duality theorem gives a recipe on how to construct a commutative algebra which can be used to study it \cite{Khalkhali2004, nLabauthors2020}. 
And given a commutative algebra, the Gelfand duality will recover the geometrical space it was constructed from. 
When commutative algebra and ordinary geometry essentially means the same, what happens if one replaces an algebra with a noncommutative one? 
This is the defining question behind \NCG, which alludes to the idea that such an algebra also describes a geometrical object, but it is very much unlike the ones from our everyday experience~\cite{Khalkhali2004, Connes1994, Connes2008}. 
In a seminal work, Bellissard et al. related the IQHE to this emerging field~\cite{Bellissard1994, Connes1994}.
When strong disorder breaks the translational symmetry of the electronic system, conventional methods of calculating the Hall conductivity become unavailable.
By formulating the problem in the language of \NCG,  Bellissard and co-workers  were able to prove that even under strong disorder, the IQHE is still given by a topological quantum number but with a different geometrical interpretation: it is now the invariant of an underlying noncommutative space and only in the case of restored transitional invariance it reduces to the well-known Chern number which characterizes the vector bundle of electronic states over the first Brillouin zone~\cite{Thouless1982, Kohmoto1985}.

At first glance, a different method in dealing with disorder in condensed matter systems is provided by gradient expansion techniques~\cite{Rammer1986,Onoda2006}.
If a perturbation is given for example by a field which varies slowly in space instead of having a constant value all throughout, clear instructions have been formulated in the past which give a systematic method of expressing physical observables as a series expansion in terms of the real-space gradients of the perturbation~\cite{Onoda2006}. 
Mathematically, the expansion is facilitated by mapping the quantum mechanical operators onto functions on a phase space which is spanned by position and momentum variables. 
The noncommutativity of the quantum operators is then encoded in a noncommutative product among the phase space functions: the Groenewold-Moyal star product~\cite{Groenewold1946, Moyal1949}.
The formal expansion of this product generates the gradient expansion.

Noncollinear magnetism presents a natural platform where the ideas of gradient expansion and noncommutative geometry could be put to use.
Smooth noncollinear spin textures are characterized by a slow variation of the local order parameter $-$ the magnetization $-$ which naturally gives rise to a parametric dependence of the Hamiltonian on local momentum and global position within the texture~\cite{Yang2011, Freimuth2013}. 
The topology of the real-space distribution of the magnetization has become a guiding principle in characterization of so-called chiral particles, such as domain walls, skyrmions or hopfions~\cite{Parkin2008,Fert2013,Sutcliffe2018,Rybakov2019}. 
Owing to their non-trivial real-space topology the properties of latter states have drawn remarkable attention motivated in part by their possible applications  as information bits of the future~\cite{Fert2017,Back2020}. 
The interplay between real-space properties of the textures with the electron transport has turned into one of the most intensively researched aspects of noncollinear magnetism, where the probe of electron dynamics modified by the presence of the noncollinearity is perceived as a unique and reliable means of texture detection and manipulation~\cite{Neubauer2009, Zeissler2018, Shao2019, Raju2019, Ahmed2019}. 

The first steps in rethinking the physics of smooth noncollinear textures in noncommutative terms have been made recently.
By resorting to the noncommutative phase space viewpoint, some of the authors have shown how to formulate the electron transport properties of smooth spin textures in terms of the noncommutative phase space star product~\cite{Lux2020}. 
Namely we demonstrated that the electric conductivity tensor is given by the so-called deformed Kubo-Bastin equations, obtained by
translating the technique of nonequilibrium Green's function of electronic quasiparticles to the setting of noncommutative phase space.
We further applied the gradient expansion technique within the two-dimensional Rashba model to single out the first order correction to the electric conductivity of spin textures.
Numerical simulations revealed that certain spin states can give rise to a large electric response, whose precise mechanism remained mysterious. 
These findings called for revisiting our knowledge of noncollinear textures based on a unified guiding principle, formulated in terms of noncommutative geometry. 

\vspace{7mm}

Here, we demonstrate that the emergent physics of noncollinear magnets can be recast into a special noncommutative type of a gauge theory. 
Namely, we show that taking the viewpoint of aperiodic noncollinear textures as a noncommutative phase space complemented by a projective module, the gradient expansions of the density of states and of the Hall coefficient are governed by a surprising connection to string theory.
This connection is characterized by an interplay between ordinary and noncommutative gauge transformations known as the Seiberg-Witten map and which was originally conceived in the effective low-energy behavior of D-branes~\cite{Seiberg1999}.
This provides a radically novel view on previous semiclassical conceptions of the  density of states and the Hall effect~\cite{Xiao2005}, while fundamentally generalizing the results of Bellissard and co-workers~\cite{Bellissard1994} to the case of metallic systems, among which the mathematical concepts of string theory could find a surprising way to practical everyday applications in the realm of noncollinear magnetism. 
We illustrate this point by revisiting the two-dimensional Rashba model and interpret its first order gradient expansion in the light of the effective Seiberg-Witten-type gauge theory.  

\vspace{7mm}

Concerning the outline of this work, the main conventions which are needed to understand the following parts of this manuscript are introduced in section \ref{sec:phase_space_formulation}. 
Section \ref{sec:nc_gauge_theory} is devoted to the formulation of the noncommutative projective module and its curvature. 
In section \ref{sec:sw_corrections}, the structure of the projective module is interpreted with the help of the Seiberg-Witten map, which is subsequently applied to the Hall effect in the Rashba model in section \ref{sec:rashba}.  
In the conclusions section we discuss the prospects of our findings and suggest possible generalizations to adjacent fields of noncollinear magnetism. 

\vfill

%----------------------------------------------------
\section{Phase space formulation}
\label{sec:phase_space_formulation}
%----------------------------------------------------

\subsection{Conventions}

Despite the fact that our analysis is based on a strictly non-relativistic setting, a relativistic notation is still useful. 
In the following, the Minkowski space will be denoted as  $\mathbb{R}^{1,d}$. 
It is the $(d+1)$-dimensional real vector space endowed with the flat metric 
$
(\eta)_{\mu\nu} \equiv \diag(-,+,+, \ldots, +).
$
With respect to these conventions, four-position and four-momentum are introduced via
$
(x)^\mu \equiv ( t, \vec{x})$ and $(p)^\mu \equiv (\epsilon, \vec{p})$ which can be summarized to a single phase space coordinate $z=(x,p)$. 
The corresponding 4-gradients on phase space are then obtained accordingly as $ \partial_{\mu} = \partial / \partial z^\mu$. 
In order to raise and lower indices on phase space, the Minkowski metric is lifted to $\spacetime\times\spacetime$ via the direct sum $\eta \to \eta \oplus \eta$.

\subsection{Wigner transformation}
\label{sec:wigner_transformation}

We employ a flavour of the Wigner transformation which can be applied to two-point functions on $\spacetime$. These are maps $o\colon\spacetime\times\spacetime\to\mathbb{C}$ which can formally be interpreted as the kernel of an operator
\begin{equation}
    \hat{o} = \int \dd x_1   \int \dd x_2 ~ o(x_1; x_2) ~ \ket{x_1} \bra{x_2},
\end{equation}
such that their Wigner transformation can be defined as~\cite{Onoda2006}
\begin{equation}
\wigner\left[\hat{o} \right](X,p) \equiv \int \dd x   \Braket{
	X + \frac{x}{2}
	|
	\hat{o}
	|
	X - \frac{x}{2}
}
e^{-i p_\mu x^\mu/\hbar} . 
\label{eq:wigner_trafo}
\end{equation} 
The Wigner transformation of an operator product is then given by
$\mathcal{W}[\hat{a}_1 \hat{a}_2  ]= \mathcal{W}[\hat{a}_1] \star \mathcal{W}[\hat{a}_2]$, where $\star$ denotes the Groenewold-Moyal star product defined by~\cite{Groenewold1946, Moyal1949, Onoda2006}
\begin{equation}
	\star \equiv \exp\left\lbrace
	\frac{i\hbar}{2} \Pi^{\mu\nu}\lpartial_\mu \rpartial_\nu
	\right\rbrace .
\end{equation}
Here,  $\Pi^{\mu\nu}= \Pi^\mu_\kappa \eta^{\kappa\nu}$ and a summation over repeated indices is implied. $\Pi$ represents the symplectic structure of the classical phase space which is given by
\begin{equation}
    (\Pi)^\mu_\nu \equiv \begin{pmatrix}
    0 & \id \\
    -\id & 0 
    \end{pmatrix}.
\end{equation}
It has the property $\Pi^2 = -\id$ and can be used to define the conjugation of phase space variables: $\Bar{z}_\mu = \Pi_{\mu\nu} ~z^\nu$ and $z^\mu = \Pi^{\nu\mu} ~\bar{z}_\nu$. Conjugation can also be applied to to the partial derivatives which gives $\bar{\partial}^\mu = \Pi^{\mu\nu}~ \partial_\nu $ and $\partial_\mu = \Pi_{\nu\mu}~ \bar{\partial}^\nu $. If usual operators on the Hilbert space such as the Hamiltonian are be transformed with $\mathcal{W}$, they first are extended to the time-domain by enforcing time-locality via the Dirac distribution $\delta(t-t')$.

The $\star$-product is the essential ingredient which turns the phase space into a noncommutative, associative, unital algebra, which we denote by $\Algebra$. 
Explicitly, it is given by
\begin{equation}
    \Algebra = \left(
    \star, C^\infty( \spacetime \times \spacetime ; \mathrm{GL}_N(\mathbb{C}) )
    \right),
\end{equation}
where $C^\infty$ denotes the space of smooth phase space functions with values in $\mathrm{GL}_N(\mathbb{C})$,  the general linear group of order $N$ over the complex numbers. 
The appearance of $\mathrm{GL}_N(\mathbb{C})$ is necessary in order to keep track of additional degrees of freedom such as the spin.
All the noncommutativity which is inherent to the position and momentum operator of quantum mechanics is now encoded into $\Algebra$.
In particular, one obtains Heisenberg's uncertainty relation $[x_i, p_i]_\star = i\hbar$.

Returning to condensed matter physics, it is usually most convenient to refer to the Green's function formalism in order to calculate statistical properties of solids ouf-of-equilibrium \cite{Rammer1986, Kita2010, Fetter2012, Abrikosov2012}. 
The so-called single-particle Keldysh Green's function on phase space, denoted by $\Gk$, can be determined from the deformed Dyson equation~\cite{Onoda2006}
\begin{equation}
    (\epsilon - H + \underline{\Sigma})\star \Gk = \id .
\end{equation}
Here, $H$ and $\underline{\Sigma}$ denote the Hamiltonian and the self-energy on phase space.
Differentiating the Dyson equation with respect to $\hbar$ leads to a differential equation of the form~\cite{Lux2020}
\begin{equation}
    \partial_\hbar \Gk =
    \frac{i}{2} \Pi^{\mu\nu} \Gk \star \partial_\mu \Gk^{-1\star} \star \Gk \star \partial_\nu \Gk^{-1\star}\star \Gk ,
\end{equation}
where $\Gk^{-1\star}$ indicates the inverse with respect to the $\star$ product. 
Other differential equations of this type will be encountered in the following sections, and they will be usually referred to as {\it flow equations}. This appeals to the idea that they describe the rate of change of quantum mechanical observables as one tunes $\hbar$ to larger values.
This differential equation can be used to construct a diagrammatic expansion of $\Gk[\Gk_0]$ in terms of a power series in $\hbar$ which is generated by the undeformed propagator $\Gk_0$ and its derivatives, given by $\Gk_0 = ( \epsilon - H + \underline{\Sigma})^{-1}$~\cite{Onoda2006, Lux2020}. 
Typically, the self-energy $\underline{\Sigma}$ is a functional of the Green's function and the solution to the Dyson equation needs to be calculated self-consistently.  
In this work, $\underline{\Sigma}$ is approximated by a broadening $\eta$, i.e., the advanced and retarded component are given by $\Sigma^\mathrm{A} = (\Sigma^\mathrm{R})^\ast =i \eta$ and the lesser component by $\Sigma^< =2 i  n_\mathrm{F}(\epsilon) \eta$, where $n_\mathrm{F}$ denotes the Fermi distribution function~with respect to~the chemical potential $\mu$. 
The limit $\eta \to 0^+$ is referred to as the clean limit. 

\subsection{Phase space eigenfunctions}
\label{sec:phase_space_eigfunc}

A central role in the phase space formulation is played by the phase space eigenfunctions whose importance has already been understood by Moyal himself and can be found in his original work~\cite{Moyal1949}. 
They form the bridge between textbook quantum mechanics, formulated in terms of Hilbert space operators, and Moyal's construction.
Let $\ket{n}$ represent an orthonormal, time-independent and complete set of states which span the Hilbert space of quantum states, i.e., 
$\sum_n \ket{n}\bra{n} = \id$ with $\braket{n | m  }  = \delta_{nm}$. Then one can construct the objects $\hat{f}_{lk} = \ket{l} \bra{k}$ and the time-local two-point functions
 $
  	\hat{f}_{lk} (x,x') \equiv \braket{\vec{x} | \hat{f}_{lk} | \vec{x}' } \delta(t-t') 
$.
One then obtains Moyal's orthogonality relation
 \begin{align}
 \int \dd \vec{z} ~ f_{lk}(\vec{z}) f_{l'k'}^\ast(\vec{z}) = h^{d} \delta_{ll'} \delta_{kk'},
 \end{align}
 which can be derived by applying the Wigner transformation to the corresponding identity on the operator level and follows then from the \emph{cyclic property} of the phase space integral:
 \begin{equation}
     	\int \dd \vec{z} ~ \phi(\vec{z})\star \psi(\vec{z}) = \int \dd \vec{z} ~ \phi(\vec{z}) \psi(\vec{z}),
 \end{equation}
 which is true for any two scalar phase space functions $\phi$ and $\psi$.
 In a similar way, one obtains the completeness relations 
$
 \int \dd \vec{z}~ f_{lk}(\vec{z}) = h^d \delta_{lk}
$ and $\sum_l f_{ll} = 1$. 
Together, these identities imply that the trace operator $\hat{\tr}$ on the Hilbert space of quantum states has an analogue on the phase space denoted by $\hat{\tr}~\hat{o} =  \Tr~\mathcal{W}[\hat{o}]$, which is given by an integration over the phase space coordinates $\vec{z}=(\vec{x},\vec{p})$. 
Explicitly, it is given by
$
     \Tr = \int~\dd\vec{z} / h^d 
$ .
If the individual functions $f_{nn}$ where indeed obtained from the quantum Hamiltonian $\hat{H}$ and correspond to an eigenvalue $\epsilon_n$, this property is inherited by the phase space Hamiltonian:
\begin{equation}
    H \star f_{nn} = \epsilon_n f_{nn},
\end{equation}
which is the defining characteristic of $f_{nn}$ as a phase space eigenfunction to the phase space Hamiltonian $H$.

\subsection{Lattice periodic systems}

An important class of phase space eigenfunctions which will be revisited in section \ref{sec:nc_gauge_theory} originates from the case of lattice periodic Hamiltonians. 
These can be diagonalized by the Bloch functions
$
	 \psi_{n\vec{k}}(\vec{x})\equiv \braket{ \vec{x} | n \vec{k} } \equiv e^{ i \vec{x} \cdot \vec{k}} u_{n \vec{k}} (\vec{x}) / \sqrt{\volume}
$ with the lattice periodic part $u_{n \vec{k}}$ and crystal volume $\volume$. 
This case is different from the previous section in the sense that periodic Born-von Karman boundary conditions need to be employed:
$
 \psi_{n\vec{k}}(\vec{x} + N^i \vec{a}_i) =  \psi_{n\vec{k}}(\vec{x}) 
$, where the vectors $\vec{a}_i$ form the basis of the Bravais lattice and the $N_i$ denote the number of unit cells in the respective direction.
Therefore, the definition of the Wigner transformation needs to be changed to~\cite{Buot1990, Iafrate2017, Culcer2006}
\begin{equation}
\wigner\left[\hat{o} \right] \equiv \int\limits_{\mathcal{V}} \dd x   \Braket{
	X + \frac{x}{2}
	|
	\hat{o}
	|
	X - \frac{x}{2}
}
e^{-i p_\mu x^\mu/\hbar},
\end{equation}
where the thermodynamic limit $\mathcal{V} \to \infty$ is taken at the end of a calculation. 
Using this recipe, the phase space eigenfunctions $f_{n\vec{k},m\vec{k}} $ can be constructed from the  $\psi_{n\vec{k}}$ by the procedure from the previous section.
For  $\vec{p} \in \mathrm{1.BZ}$ (the first Brillouin zone), an explicit calculation shows that~\cite{Iafrate2017}
\begin{align}
f_{n\vec{k},m\vec{k}} (\vec{x},\vec{p}) = \delta_{\vec{p}/ \hbar, \vec{k} }
\sum_{ \vec{G}  } 
~\hat{u}_{n\vec{p}/\hbar} (\vec{G}) \hat{u}^\ast_{m\vec{p}/\hbar} (-\vec{G})e^{2i \vec{x} \cdot  \vec{G}},
\end{align}
where $\vec{G}$ represents the reciprocal lattice vectors.
Moyal's phase space coordinate $\vec{p}$ can therefore be identified with the crystal momentum $\hbar \vec{k}$. 
Let $G_0$ represent the phase space Green's function of a lattice periodic system and $V$ a slowly varying perturbation, breaking the translational symmetry. 
By virtue of the Dyson equation, the renormalized Green's function is given by $G = G_0 + G_0 \star V \star G$ and due to the slowly varying nature of the perturbation, it can still be approximately expanded in the Bloch basis:
 \begin{equation}
 G(\vec{x}, \vec{p}) \approx \sum_{nm } \sum_{\vec{k} \in 1. \mathrm{BZ} } g^{nm}_{ \vec{k}}(\vec{x}) f_{n\vec{k},m\vec{k}}~ (\vec{x}, \vec{p}),
 \end{equation}
If one considers the general class of observables $o$ with slowly varying kernels $o^{nm}_{\vec{k}}(\vec{x})$ that are constant over the range of one unit cell, but change in the course of many, one finds
  \begin{align}
& \frac{1}{h^d} \int\dd \vec{x} \dd \vec{p} \sum_{\vec{k} \in 1. \mathrm{BZ} }
 \sum_{nm} o^{nm}_{\vec{k}}(\vec{x}) f_{n\vec{k}m\vec{k}} ( \vec{x}, \vec{p})
 \notag \\ &
 = \frac{1}{\mathcal{V}} \int \dd\vec{r}\sum_{\vec{k} \in 1. \mathrm{BZ} } \tr ~ o_\vec{k} (\vec{r}) ,
 \end{align}
 where the trace $\tr$ sums over the band indices. This result can be proven by subdividing the domain into unit cells and then translating the summation over the lattice into a Riemann sum. The zeroth order gradient expansion of the expectation value of a $\star$-product of slowly varying operators is therefore given by the correspondence
\begin{equation}
	\Tr \lbrace O^1 \star \ldots \star O^n \rbrace
	~\leftrightarrow~
	\frac{1}{\mathcal{V}} \int \dd\vec{r}\sum_{\vec{k} \in 1. \mathrm{BZ} } \tr ~ O^1_\vec{k}(\vec{r}) \ldots O^n_\vec{k}(\vec{r}),
	\label{eq:correspondence}
\end{equation}
which can be viewed as the leading term of a more general gradient expansion.

%----------------------------------------------------
\section{Noncommutative gauge theory}
\label{sec:nc_gauge_theory}
%----------------------------------------------------

\subsection{Anomalous Hall effect}

After having introduced the necessary mathematical language, we now want to turn to the main results of our manuscript.
In a recently published work, we have derived a deformed version of the Kubo-Bastin equations~\cite{Lux2020}. 
These equations express the electrical conductivity tensor  of a spin texture (or generally non-periodic solid) in terms of the Green's function defined on the phase space.
In the particular case of two dimensional systems, the off-diagonal elements of the conductivity tensor are determined by 
\begin{align}
    \sigma^\mathrm{sea}_{xy} &= \hbar e^2\Re ~
    \fraktr_\mathrm{sea} \Braket{  \Gret \star v_x \star \Gret  \star v_y \star \Gret - ( x \! \leftrightarrow \! y) }, \label{eq:deformed_kubo1}
    \\
     \sigma^\mathrm{surf}_{xy} &= \hbar e^2  \Re ~
    \fraktr_\mathrm{surf} \Braket{
    v_x \star ( \Gret - \Gadv ) \star v_y \star \Gadv 
    }, \label{eq:deformed_kubo2}
    \notag \\
\end{align}
where the bracket $\braket{\bullet}$ indicates the average over the real-space coordinates and the phase space trace is extended as to integrate also over energy:
\begin{equation}
    \fraktr ( \bullet ) \equiv   \frac{1}{2\pi}  \int \dd\epsilon
    \int \frac{ \dd^2 \vec{p}}{(2 \pi \hbar )^{2}}~\tr ( \bullet ) .
\end{equation}
The index $\fraktr_\mathrm{sea}$ indicates that the energy integral is to be weighted with the Fermi distribution function $n_\mathrm{F}$, while in $\fraktr_\mathrm{surf}$ it is to be weighted with respect to its first derivative $\dd n_\mathrm{F} / \dd \epsilon$. 
The net conductivity is then given by the sum $\sigma_{xy}=\sigma^\mathrm{sea}_{xy}+\sigma^\mathrm{surf}_{xy}$ and relates the electric response current $j_x$ to an applied electric field $E_y$. 

In lattice periodic systems, it is known that for insulators and for clean metals, the Kubo-Bastin equations should reduce to a purely geometrical construction at zero temperature. 
The anomalous Hall conductivity is then given by a momentum space integral over the Berry curvature of the occupied subspace~\cite{Jungwirth2002, Nagaosa2010}.
While this result was generalized by Bellissard to the case of aperiodic insulators where the Fermi energy $\mu$ belongs to a gap of extended states~\cite{Bellissard1994}, no such generalization has been known to exist for metals.

Here, we propose that Bellissard's results can be obtained directly from the phase space formulation and can even be generalized to metallic systems.  
Starting point is the phase space density matrix defined by
\begin{equation}
    \rho \equiv  - \frac{1}{\pi}\int\limits_{-\infty}^{+\infty} \dd \epsilon ~  n_\mathrm{F}(\epsilon) ~\Im~\tr~\Gret .
    \label{eq:dens_mat}
\end{equation}
In the limit of zero temperature and in the absence of scattering, the density matrix describes a pure state, which means that $\rho \star \rho = \rho$ (projector condition).
For example, if the phase space eigenfunctions $f_{nn}$ as defined in section \ref{sec:phase_space_eigfunc} are known, the zero temperature density matrix of an insulator is simply given by the projector $\rho\equiv  \sum_{\epsilon_n <  \mu} f_{nn} $.

By performing the energy integrations in Eq.~(\ref{eq:deformed_kubo1}) and Eq.~(\ref{eq:deformed_kubo2}) exactly (see appendix~\ref{app:projector_formula}), we arrive at the first central result of this work, which is given by the anomalous Hall coefficient
\begin{equation}
    \sigma_{xy}  =   \frac{e^2}{h}~  \frac{1}{2\pi } \int \dd^2\vec{p}~ \Omega_{p_x p_y}(\vec{p}).
    \label{eq:ahe_coeff}
\end{equation}
In analogy to the ordinary formulations of $\sigma_{xy}$ in a lattice periodic systems, one can identify the noncommutative analogue of the Berry curvature  $\Omega_{p_x p_y}(\vec{k}) $, which is given by the expression
\begin{equation}
    \Omega_{p_x p_y}(\vec{k}) \equiv i \tr \Braket{ 
          \rho \star [ \partial_{p_x} \rho, \partial_{p_y} \rho]_\star
         } .
    \label{eq:berry}
\end{equation}
If the Hamiltonian has translational symmetry, the correspondence principle from Eq.~(\ref{eq:correspondence}) maps this result directly to the known expressions for lattice periodic systems~\cite{Thouless1982, Kohmoto1985, Jungwirth2002, Xiao2010, Nagaosa2010}.
And while the result is exact only in the clean limit for $T \to 0$ where the projector condition is fulfilled, it can be expected to remain a good approximation at small, but finite temperatures.
The importance of this result is that it facilitates a gradient expansion for clean metals in purely geometrical terms, as an expansion of the Berry curvature (by expansion of the $\star$-products).
While such an expansion can of course be done already on the level of Eq.~(\ref{eq:deformed_kubo1}) and Eq.~(\ref{eq:deformed_kubo2}), it would be much harder to unveil this geometric interpretation.
However, Eq.~(\ref{eq:berry}) has been identified as ``curvature" so far only in an analogy to the ordinary formulation. 
In the following section, we will explore how to make this statement precise by interpreting it as the true curvature of a noncommutative fiber bundle.

\subsection{Noncommutative fiber bundle}

While in the lattice periodic case, the interpretation of the anomalous Hall effect in terms of a Bloch bundle is well-known~\cite{Thouless1982, Kohmoto1985, Haldane1988, Jungwirth2002, Panati2007}, this interpretation is better thought of as an analogy in the noncommutative realm.
The Serre-Swan theorem establishes a correspondence between vector bundles and projective modules on \emph{commutative} rings~\cite{Serre1955, Swan1962}. 
Fig.~\ref{fig:bloch_bundle} illustrates this correspondence for the ordinary case of a one dimensional Brillouin zone which has the shape of a circle. 
In its essence, the occupied states at each $\vec{k}$-point define a vector space which is a subspace of the full Hilbert space. 
The collection of these local vector spaces parameterized by Brillouin zone is what forms the vector bundle~\cite{Frankel2011, Nakahara2003} and gives the Bloch bundle its name.
An assignment of these vector spaces is however equivalent to the assignment of a projection operator $\varrho_\vec{k}$ to each $\vec{k}$-point, whose image is spanned by the occupied states.
And since projection operators are essentially an algebraic construction, they are well-suited for a generalization to noncommutative spaces, in which the underlying algebra is simply replaced by a noncommutative one~\cite{Connes1994,Khalkhali2004, Masson2012}. 
A natural projective module $M$ in the context of the noncommutative quantum Hall effect is given by
\begin{equation}
    M = \rho \star \Algebra \star \rho ,
\end{equation}
where $\rho$ is again the density matrix projector from the previous section and $\Algebra$ is the algebra of phase space functions (see section \ref{sec:wigner_transformation}).
The module structure is established by the left and right $\star$-action of the gauge group
\begin{equation} 
    \mathcal{G}(M)
    =\lbrace
    U \in M ~|~ U^\dagger \star U =U \star U^\dagger = \id
    \rbrace,
\end{equation}
which represents a unitary change of frame in the algebra $\Algebra$. 
In order to discuss the geometric properties of $M$, we will follow closely the discussion of~\cite{Masson2012}.
A first ingredient are the tangent vectors of $\Algebra$, whose role is played by the space of derivations $\mathrm{Der}(\Algebra)$, consisting of linear maps $\mathfrak{X}: \Algebra \to \Algebra$ and which obey they Leibniz rule  $\mathfrak{X}(a\star b) = \mathfrak{X}(a) \star b + a \star \mathfrak{X}(b)$. 
In differential geometry, this is a standard way to construct the tangent space $\mathrm{T}N$ to a manifold $N$: as derivations on the algebra of smooth functions $\mathrm{Der}(C^\infty(N)) \cong \Gamma (\mathrm{T}N)$~\cite{Lee2013}. 
In the case of $\Algebra$, all derivations $\mathfrak{X}$ are so-called inner derivations~\cite{Masson2012}, which means that there always exists an $x \in \Algebra$ such that $\mathfrak{X}(a) = [x,a]_\star \equiv \ad^\star_x ~a$. 
In particular, the partial derivatives applied to elements of $\Algebra$ act as derivations. 
The correspondence is established by the adjoint of the conjugate variable: $\partial_\mu = - i \ad_{ \bar{z}_\mu} ^\star \in\mathrm{Der}(\Algebra)$. 

In ordinary differential geometry, the application of a tangent vector to a function yields a directional derivative. 
Drawing an analogy, for $\psi \in \Algebra$, $ \mathfrak{X}(\psi)$ can be interpreted as a quantum derivative of $\psi$~\cite{Connes1985, Connes1994}.
In general, this derivative will have components which do not belong to the projective module anymore. 
This leads to the concept of covariant differentiation, defined as
\begin{equation}
    \cov_{\mathfrak{X}}  \psi \equiv \rho \star \mathfrak{X}(\psi) \star \rho,
\end{equation}
which projects the derivative back into the module.
It fulfills the parallel transport condition $\cov_{\mathfrak{X}} \rho = 0$ for all $\mathfrak{X}\in\mathrm{Der}(\Algebra)$. 
If $\mathfrak{X}$ is generated by an element $x \in M$ as $\mathfrak{X}=\ad^\star_x$ then $\cov_{\mathfrak{X}} \psi = 0$ for all $\psi \in M$. 
This corresponds to intuition that the covariant derivative has no components sticking out of the tangent space. 
For the direction of $\mathfrak{X}=\partial_\mu$, the covariant derivative can also be written as
\begin{equation}
    \cov_\mu = \partial_\mu - i \mathfrak{A}_\mu ,
\end{equation}
where the connection coefficients $ \mathfrak{A}_\mu$ are derivations
\begin{equation}
    \mathfrak{A}_\mu ( \psi )  \equiv  \ad^\star_{ \Conn_\mu } \psi \equiv [ \underbrace{-i [\partial_\mu \rho, \rho]}_{\Conn_\mu}  ,\psi]_\star .
\end{equation}
Since the $ \mathfrak{A}_\mu$'s and their generators $\Conn_\mu$ correspond to the covariant derivative which fulfills the parallel transport condition $\nabla_\mu \rho = 0$, we also refer to this construction as the \emph{parallel transport connection}.

\figureI

Now that the connection is specified, the noncommutative generalization of the Riemannian curvature can be defined as~\cite{Masson2012} 
\begin{equation}
    R_\star(\mathfrak{X},\mathfrak{Y})  = i  [ \nabla_\mathfrak{X}, \nabla_\mathfrak{Y}]_\star \star   - i \nabla_{[ \mathfrak{X}, \mathfrak{Y}]}  .
\end{equation}
When evaluated on the coordinate derivations, one finds
\begin{equation}
   R_\star(\mathfrak{X}_\mu,\mathfrak{X}_\nu)  = \partial_{\mu} \mathfrak{A}_\nu - \partial_{\nu} \mathfrak{A}_\mu - i [\mathfrak{A}_\mu , \mathfrak{A}_\nu]_\star \equiv \mathfrak{F}_{\mu\nu},
\end{equation}
where one can identify the adjoint  $\mathfrak{F}_{\mu\nu} = \ad_\star ~ \Curv_{\mu\nu}$ with
\begin{equation}
    \Curv_{\mu\nu} = \partial_\mu \Conn_\nu - \partial_\nu \Conn_\mu - i [\Conn_\mu, \Conn_\nu]_\star.
\end{equation}
For the parallel transport connection this evaluates to 
\begin{equation}
    \Curv_{\mu\nu} =  i  [ \partial_{\mu} \rho, \partial_{\nu} \rho]_\star .
\end{equation}
This establishes the desired result: the analogue of the Berry curvature which has been identified in Eq.~(\ref{eq:berry}) is really the Riemann curvature of the projective module defined by the projection operator $\rho$:
\begin{equation}
    \Omega_{\mu\nu} = \tr~ \rho \star \Curv_{\mu\nu} .
\end{equation}
It is worth to reflect the meaning of this result, since it is tempting to try to find some visual interpretation similar to that for the classical case which was depicted in Fig.~\ref{fig:bloch_bundle}.
This is however not possible due to the way which has been taken in order to arrive at this result.
Noncommutative geometry fundamentally operates on analogies.
In this case, it was based on the duality between vector bundles and projective modules over commutative rings. 
Subsequently, the commutative ring has been replaced with a noncommutative ring, but still it is being viewed as a representative for some noncommutative version of a vector bundle.
It is therefore an object which is very remote from all the geometric spaces of our everyday experience: one cannot draw a simple picture, but certain geometrical properties like the curvature can still be defined in a meaningful way.

\subsection{Gauge covariance}

To elucidate the role of the module structure and the space $  \mathcal{G}(M)$, we want to briefly adress the gauge transformations in $M$.
A gauge transformation in the projective module $M$ corresponds to a unitary change of frame in $\Algebra$.
Explicitly, we want to consider gauge transformations of objects $\psi = \rho \star \psi \star \rho\in M$. 
Under a $\star$-unitary transformation $U\in\mathcal{G}(M) $ these objects change as
\begin{equation}
    \psi \to U \star \psi \star U^\dagger = \psi' .
    \label{eq:adjoint_representation}
\end{equation}
The covariant derivative of $\psi$ should change covariantly by definition, which dictates the behavior of $\Conn_\mu$ under gauge transformations:
\begin{equation}
    \cov_\mu \psi = \cov_\mu \big( \underbrace{U^\dagger \star  \psi' \star U}_{\psi} \big)
    \overset{!}{=} U^\dagger \star \cov_\mu' \psi'\star U.
\end{equation}
From the explicit evaluation of this conditions one finds the familiar transformation behavior~\cite{Nakahara2003}
\begin{equation}
    \Conn_\mu \to U \star\Conn_\mu \star U^\dagger - i \partial_\mu U \star U^\dagger = \Conn_\mu' .
\end{equation}
The coefficients of the connection are therefore not gauge-covariant. For the curvature on the other hand, one finds
\begin{equation}
    \Curv_{\mu\nu} \to U \star \Curv_{\mu\nu} \star U^\dagger = \Curv_{\mu\nu}' .
\end{equation}
This can be seen by either inserting the transformation behavior of $\Conn_\mu$ directly, or by first rewriting the curvature in terms of derivations, i.e.,
\begin{equation}
   \Curv_{\mu\nu} = i [ \mathfrak{X}_\mu(\rho), \mathfrak{X}_\nu(\rho)]_\star .
\end{equation}
Since all derivations on $\Algebra$ are inner derivations, $\forall \mathfrak{X}\in \mathrm{Der}(\Algebra): \exists x \in \Algebra:  \mathfrak{X} = \ad_\star x $, one has
\begin{equation}
    U \star \mathfrak{X} (\psi) \star U^\dagger =  \mathfrak{X}'_\mu (\psi') .
\end{equation}
Here we denote 
$
    \mathfrak{X}'_\mu = \ad_\star( U\star x \star U^\dagger )
$ and $\psi, \psi'$ as above. Therefore, the curvature transforms covariantly.  In full analogy to the construction of the curvature, one can define a noncommutative metric as the map
\begin{align}
  &  \Qmetric\colon \mathrm{Der}(\Algebra) \times \mathrm{Der}(\Algebra)  \to \Algebra
  \\
  & ( \mathfrak{X}, \mathfrak{Y} ) \mapsto \Qmetric( \mathfrak{X}, \mathfrak{Y})  \equiv  \lbrace \mathfrak{X}(\rho) , \mathfrak{Y}(\rho) \rbrace_\star .
\end{align}
This means that the noncommutative metric is a gauge-covariant object, just as the curvature. Further, it is symmetric: $\Qmetric( \mathfrak{X}, \mathfrak{Y}) = \Qmetric( \mathfrak{Y}, \mathfrak{X}) $ and positive-definite, i.e., $\Qmetric( \mathfrak{X}, \mathfrak{X})>0$ if $\mathfrak{X}(\rho)>0$. When evaluated on the coordinate derivations one finds
\begin{equation}
    \Qmetric( \mathfrak{X}_\mu, \mathfrak{X}_\nu) = \lbrace \partial_\mu \rho, \partial_\nu \rho \rbrace_\star \equiv \Qmetric_{\mu\nu} ,
    \label{eq:quantum_metric_tensor}
\end{equation}
and is therefore very similar to previous conceptions of a quantum metric tensor~\cite{Ma2010}.

\subsection{Relation to the work of Bellissard}
\label{sec:relation_to_bellissard}

Eq.~(\ref{eq:ahe_coeff}) resembles very much the result of Bellissard and co-workers~\cite{Bellissard1994}, but without explicitly invoking the construction of what he refers to as the \emph{noncommutative Brillouin zone}~\cite{Bellissard1994}, and which he has used to relate the Chern character to the index of a Fredholm operator, thus verifying the quantization and integrality of the Hall conductance.
Nevertheless, one can see that the phase space construction is actually equivalent to Bellissard's original idea.
In a system with broken translational invariance, the noncommutative Brillouin zone is thought of as the virtual object over which the family of noncommuting translates of the Hamiltonian exists. 
This family generates a noncommutative algebra of observables over this object, accessible to Connes' tools from noncommutative geometry.
%This family is a noncommutative algebra and, therefore, accessible to Connes' tools from noncommutative geometry.
To endow it with the structure of a noncommutative manifold, Bellissard imposes rules for calculus. 
The analogue to the integration over the Brillouin zone is given by a trace per unit volume, coinciding with the phase space trace we introduced in our formalism. 
%Also, a noncommutative counterpart to the differentiation with respect to the crystal momentum, acting in his representation over Hilbert space, is now defined as a quantum differential, i.e., the commutator with the position operator. 
And a noncommutative counterpart to the differentiation with respect to the crystal momentum is now defined as a quantum differential, i.e., the commutator with the position operator. 
This corresponds to our inner derivation definition of the partial derivatives.

The essential difference to the approach of Bellissard is the application of the Wigner transformation, Eq.~(\ref{eq:wigner_trafo}).
Some of the abstract concepts such as the quantum differentials actually return to their intuitive meanings in the transition to the phase space formulation of quantum mechanics (not only by analogy). 
For example, the $\star$-commutator of phase space functions with the position operator is by construction  formally identical to  their momentum derivative. 
Further, the phase space formulation facilitates a study of the semiclassical limit and the gradient expansion, which will be the subject of the following section.

%Bellissard proceeds to derive the Kubo formula from electronic transport theory and relates it to the Chern character of the Fermi projection in the above algebra. Further, he relates the Chern character to the index of a Fredholm operator. This verifies the quantization and integrality of the Hall conductance.

%----------------------------------------------------
\section{Seiberg-Witten Corrections}
\label{sec:sw_corrections}
%----------------------------------------------------

In condensed matter physics, the phase space formulation is regularly invoked in order to calculate gradient corrections~\cite{Rammer1986, Onoda2006}.
This is particularly interesting for noncollinear spin configurations in real space which break the translational symmetry of the underlying crystal lattice.
Physical properties such as the conductivity can then be formulated as a tensor field which couples to the gradients of magnetization texture, but which themselves only require the knowledge of the eigenstates in the collinear configuration, which can be a great computational benefit.
In section \ref{sec:nc_gauge_theory}, we have introduced a phase space formulation of the density matrix and the Berry curvature and therefore, there is in principle the possibility to investigate their gradient expansions.

\subsection{Correction of the projection operator}

In order to determine the gradient  corrections to the Berry curvature, one needs to determine the corrections to the density matrix first. 
At zero temperature and in the clean limit, the density matrix $\rho$ fulfills
$
    \rho \star \rho = \rho 
$, and is therefore a projection operator on phase space.
This requirement is very restrictive. 
So restrictive in fact, that it almost uniquely determines the gradient expansion of $\rho$. 
Differentiating this constituting equation with respect to $\hbar$ yields
\begin{equation}
    \partial_\hbar \rho = \lbrace \partial_\hbar \rho , \rho \rbrace_\star + \frac{i}{2} \Pi^{\mu\nu}  \partial_\mu \rho \star  \partial_\nu \rho .
    \label{eq:sw_projector_a}
\end{equation}
Essentially, this is a differential equation for the projection operator with an initial condition $\rho \to \varrho$ for $\hbar \to 0$. A way to construct $\varrho$ is given by the $\hbar\to 0 $ limit of Eq.~(\ref{eq:dens_mat}). 
Solving Eq.~(\ref{eq:sw_projector_a}) for the derivative term gives (see appendix \ref{app:projection_flow} for a derivation)
 \begin{align}
    \partial_\hbar \rho     =   - \frac{1}{4} \Pi^{\kappa\lambda} \lbrace \Conn_\kappa, \partial_\lambda \rho  + \nabla_\lambda \rho \rbrace_\star  
     \label{eq:sw_projector_b} .
\end{align}
However, arbitrary interband terms could be added to Eq.~(\ref{eq:sw_projector_b}) while leaving Eq.~(\ref{eq:sw_projector_a}) invariant. 
For any operator $o$, we understand the interband terms as the projections $\rho \star o \star \rho^\perp$ and $\rho^\perp \star o \star \rho$, where $\rho^\perp = \id - \rho$. 
For insulators, an alternative way to construct the same equation can be obtained from the Green's function using the deformed residue formula
\begin{equation}
    \rho = \frac{1}{2\pi i } \oint\limits_\gamma ~\dd z~ (z - H)^{-1\star},
\end{equation}
where the complex contour $\gamma$ encloses the real energy axis anit-clockwise up to the Fermi energy. 
The limit $\hbar\to 0$ of this integral determines $\varrho$. The result for $\partial_\hbar \rho$ is in agreement with Eq.~(\ref{eq:sw_projector_b}). The interband terms are shown to decay as $1/\Delta E$ where $\Delta E$ is the size of the band gap. Eq.~(\ref{eq:sw_projector_b}) is therefore quite remarkable, since it gives an iterative procedure for the full projection operator in an insulating system in the adiabatic limit without performing any contour integration. 
In the following, we neglect the complications due to the interband terms and consider implicitly the adiabatic limit $\Delta E \to \infty$. 
An interesting consequence of Eq.~(\ref{eq:sw_projector_b}) is given by the projection onto the occupied subspace:
\begin{equation}
    \rho \star \partial_\hbar \rho \star \rho  = - \frac{1}{4} \Pi^{ij} \rho\star\Curv_{ij} .
\end{equation}
This result can be interpreted as the change of the phase space volume which is associated with variation of $\hbar$ and can be elegantly obtained from the $\star$-product formalism.  
If one now takes an effective two-level system, the zeroth order term has trace one, i.e., $\tr~ \varrho = 1$. 
Defining the phase space density of states as $\mathcal{D} = \tr ~\rho / (2\pi)^d$, one therefore finds
\begin{align}
\mathcal{D}
&= \frac{1}{(2 \pi)^d} \left( 1 +  \sum_i \left. \Omega_{k_i x_i} \right|_{\hbar \to 0} +  \mathcal{O}(\hbar^2)\right) ,
\label{eq:niu_dos}
\end{align}
where we have again used the notation:
$
   \Omega_{\mu\nu}   = \tr~\rho \star  \Curv_{\mu\nu}
$. 
This is exactly the result which was obtained by Xiao et al. by referring to the semiclassical wave packet formalism~\cite{Xiao2005, Xiao2010}.
Here it appears as a necessary consequence from the restrictions which are imposed by the algebraic condition $\rho\star \rho = \rho$.

\subsection{St\v{r}eda formula}

The expectation value of the electron density $n_\mathrm{e}$ is given by the $d$-dimensional phase space integral over $\mathcal{D}$, i.e., 
\begin{equation}
    n_\mathrm{e}
    = \frac{1}{V}\int \dd^d \vec{k}\dd^d \vec{x} ~ \mathcal{D} .
    \label{eq:particle_dense}
\end{equation}
In a thought experiment, one can imagine that the system is permeated by a magnetic field $B$ through  the $xy$ plane. This modifies the symplectic structure of the phase space and the $\star$ product changes accordingly:
\begin{equation}
	\star \equiv \exp\left\lbrace
	\frac{i\hbar}{2} ( \Pi^{\mu\nu}\lpartial_\mu \rpartial_\nu - e B \epsilon^{ijz} \lpartial_{p_i} \rpartial_{p_j} )
	\right\rbrace .
\end{equation}
At the same time, the alteration of the $\star$ product leads to a shift of the projection operator which can be obtained with the same technique as before:
\begin{equation}
    \partial_B \rho = \lbrace
    \partial_B \rho, \rho \rbrace_\star - \frac{ie\hbar}{2} [ \partial_{p_x}\rho , \partial_{p_y} \rho]_\star .
\end{equation}
The cyclic property of the phase space trace eliminates possible interband corrections to $n_\mathrm{e}$, such that one can focus on the occupied subspace. A projection unveils the deformed momentum space curvature 
\begin{align}
   \rho \star \partial_B \rho \star \rho 
   =  \frac{e \hbar}{2}  ~  \rho \star \Curv_{p_x p_y}.
\end{align}
Applying this relation to the electron density in Eq.~(\ref{eq:particle_dense}), one only needs to compare the result with the Berry curvature expression for the Hall effect in Eq.~(\ref{eq:ahe_coeff}) to establish the noncommutative generalization of the St\v{r}eda formula~\cite{Streda1982}: 
\begin{equation}
    \sigma_{xy} = e \frac{\partial n_\mathrm{e}}{ \partial B}  .
    \label{eq:streda}
\end{equation} 
This equation is valid for even for aperiodic systems, which has been established in the past by referring to noncommutative geometry~\cite{Rammal1990, Prodan2016}. 

\subsection{Seiberg-Witten map}

Unexpectedly for the perspective of condensed matter theory, the differential equation which governs the projector in Eq.~(\ref{eq:sw_projector_b}) is known from a rather unusual context: effective field theories in string theory. 
In studying the low-energy behaviour of open strings, Seiberg and  Witten encountered a problem: different regularization schemes would provide them with different gauge theories. One commutative, one noncommutative~\cite{Seiberg1999}. 
They concluded that there should exist a map among commutative and noncommutative gauge fields in order to have a consistent physical description, independent of the regularization scheme which is employed. 
We are now facing an analogous situation which becomes evident in case of lattice periodic systems: either one employs the ordinary gauge theory on the Bloch bundle or the noncommutative gauge theory in the phase space geometry. 
Both should give same answer. 
This means there exists a map between them and this map can be used to study gradient corrections in a systematic way.

The construction of the Seiberg-Witten map is best understood from the interplay of ordinary and noncommutative infinitesimal gauge transformations. 
In our discussion, we follow closely the results of Jurc\v{o}  and co-workers \cite{Jurco2000, Jurco2001}.
Let $U \in \mathcal{G}(M)$ be close to the identity such that one can expand
$
    U \approx \id + i \Lambda 
$, where $\Lambda \in \Algebra$ is a hermitian gauge parameter, which depends on the phase space coordinates.
Under an infinitesimal gauge transformation of this form, an object $\psi \in M$ changes by an amount
\begin{equation}
    \delta^\star_\Lambda \psi = i~ \ad_\Lambda^\star \psi .
\end{equation}
 From this one obtains the relation
\begin{equation}
    [ \delta_\Lambda^\star , \delta_\Gamma^\star]\psi= \delta_{-i [ \Lambda, \Gamma]_\star}^\star  \psi.
\end{equation}
In ordinary gauge theory, the gauge parameters $\Lambda$ take values in a Lie algebra $\mathfrak{g}$ which has a finite number of generators $T^a$.
Every Lie algebra  is closed under its Lie bracket which for matrix Lie algebras is just the usual commutator.
For ordinary gauge fields in absence of $\star$-product, the commutator of infinitesimal gauge transformations is therefore again generated by a Lie algebra valued gauge parameter.
The algebra of gauge transformations thus respects the structure of the underlying Lie algebra: it is a Lie algebra homomorphism.

For noncommutative gauge transformations, this does not seem to be the case at first.
If one assumes an expansion of the gauge parameters $\Lambda = \Lambda_a T^a$ and $\Gamma = \Gamma_a T^a$ in terms of generators $T^a$ of $\mathfrak{g}$, their $\star$-commutator is given by
\begin{equation}
    [\Lambda, \Gamma]_\star = \frac{1}{2} \left( 
    [\Lambda_a, \Gamma_b]_\star \lbrace T^a, T^b \rbrace
+   \lbrace \Lambda_a, \Gamma_b \rbrace_\star [ T^a, T^b ]
    \right) . 
\end{equation}
Since $\lbrace T^a, T^b \rbrace$ is in general not Lie algebra valued anymore, the commutator of gauge transformations closes only within the universal enveloping algebra $U(\mathfrak{g})$. Any $\Lambda \in U(\mathfrak{g})$ then has the general expansion
\begin{equation}
    \Lambda = \sum_{n=1}^\infty \Lambda^{n-1}_{a_1\cdots a_n} :T^{a_1} \cdots T^{a_n}:  ,
    \label{eq:envelope_expansion}
\end{equation}
where the colons indicate the symmetrization over the symmetric group $S_n$:
\begin{equation}
    :T^{a_1} \cdots T^{a_n}: ~\equiv~ \frac{1}{n!} \sum_{\pi \in S_n} T^{a_{\pi(1)}} \cdots T^{a_{\pi(n)}} .
\end{equation}
This construction now incorporates all possible combinations of Lie algebra generators.

Unfortunately, the situation seems intractable now that an infinite number of gauge parameters had to be introduced~\cite{Jurco2001}. 
The problem can be fixed by a physical insight which is especially lucid in the context of the gradient expansion.
Generally speaking, a gradient expansion gives a local approximation to the electronic structure.
It therefore also carries information about the local, ordinary gauge degrees of freedom.
If one is able to diagonalize the exact quantum mechanical Hamiltonian at the same time, one would also have access to the Wigner transformed gauge theory of the full system, thus representing the noncommutative counterpart.
Intuitively speaking, an ordinary gauge transformation which is applied to the local electronic structure everywhere should amount to a consistent noncommutative gauge transformation applied to the global electronic structure:
\begin{align}
    \begin{pmatrix}
    \text{ordinary} \\
    \text{local}
    \end{pmatrix}
    \leftrightarrow
    \begin{pmatrix}
    \text{noncommutative} \\
    \text{global}
    \end{pmatrix}.
\end{align}
Requiring the consistency between these two different viewpoints is what tames the infinite number of gauge parameters in the universal enveloping algebra.

Mathematically, this synthesis is achieved by redefining the noncommutative gauge parameters in harmony with their commutative counterparts~\cite{Jurco2001}, i.e.,
\begin{align}
     \delta_\alpha \psi_0 & = i \ad_\alpha~ \psi_0
     \\
     \delta_\alpha^\star \psi  &\equiv i \ad_{\Lambda_\alpha}^\star~ \psi,
\end{align}
where the first equation is the $\hbar \to 0$ limit of the second and where $\alpha = \left.\Lambda_{\alpha}\right|_{\hbar\to 0} $.
The noncommutative gauge parameters $\Lambda_\alpha$ can then be constructed in such a way that the corresponding infinitesimal $\star$-gauge transformations respect the Lie algebra structure of the ordinary, local generators~\cite{Jurco2001}:
\begin{equation}
    [ \delta_\alpha^\star , \delta_\beta^\star] \rho \overset{!}{=} \delta_{-i [\alpha, \beta]}^\star ~ \rho .
\end{equation}
The differentiation with respect to $\hbar$ yields a differential equation and a class of inhomogenous solutions is represented by the flow equation~\cite{Jurco2001}
\begin{equation}
    \partial_\hbar \Lambda_\alpha [\conn] = \frac{1}{4} \Pi^{\mu\nu} \lbrace \partial_\mu \Lambda_\alpha, \Conn_\nu \rbrace_\star.
\label{eq:gauge_consistency_flow}
\end{equation}
This solution is however not unique and arbitrary homogeneous solutions can be added~\cite{Asakawa1999}. 
Once the gauge parameter is fixed, the Seiberg-Witten map for the noncommutative connection $\Conn$ can be obtained from the relation~\cite{Seiberg1999}
\begin{equation}
    \Conn [ \conn + \delta_\alpha \conn ]
    =
    \Conn [\conn] + \delta_{\alpha}^\star \Conn [\conn],
\end{equation}
where $\conn$ is the $\hbar\to 0$ limit of $\Conn$ and thus represents the ordinary, local connection.
The philosophy of the Seiberg-Witten map can therefore be summarized in the commutative diagram~\cite{Asakawa1999}
\begin{equation}
\begin{tikzcd}
\conn_\mu \arrow[d,swap,"\delta_\alpha"] \arrow[r] & \Conn_\mu \arrow[d,"\delta_{\alpha}^\star"] \\
\conn_\mu' \arrow[ r] & \Conn_\mu' 
\end{tikzcd}
\end{equation}
This time, it means that the behavior of $\Conn$ under noncommutative gauge transformations should be consistent with the behavior of $\conn$ under ordinary gauge transformations.
Differentiation by $\hbar$ yields a flow equation for connection. 
Using the solution from the gauge consistency criterion in Eq.~(\ref{eq:gauge_consistency_flow}) results in
\begin{equation}
    \partial_\hbar  \Conn_\mu = - \frac{1}{4} \Pi^{\kappa\lambda}
    \lbrace
    \Conn_\kappa , \partial_\lambda \Conn_\mu
    + \Curv_{\lambda\mu}
    \rbrace_\star .
    \label{eq:sw_eq_A}
\end{equation}
 Once the Seiberg-Witten map for $\Conn$ is known, it can be used to compute the corresponding map for the curvature $\Curv$. The homogenous solutions are given by
\begin{align}
    \partial_\hbar \Curv_{\mu\nu}
    & = - \frac{1}{4} \Pi^{\kappa\lambda} \big( \lbrace \Conn_\kappa, \partial_\lambda \Curv_{\mu\nu} + \nabla_\lambda \Curv_{\mu\nu} \rbrace_\star
   \notag\\ & \hspace{3.5cm}
     -2 \lbrace \Curv_{\mu\kappa} , \Curv_{\nu\lambda} \rbrace_\star\big) . 
     \label{eq:sw_curvature}
\end{align}
A Seiberg-Witten map can also be written down for matter fields which transform in the adjoint representation~\cite{Uelker2008}.  
These are fields, where the gauge transformations act from the left and from the right simultaneously and which is exactly the case for the projective module we constructed in section \ref{sec:nc_gauge_theory} as can explicitly be seen in Eq.~(\ref{eq:adjoint_representation}).
As shown in~\cite{Uelker2008}, the Seiberg-Witten map associated with a matter field in the adjoint representation agrees precisely with the flow behavior in Eq.~(\ref{eq:sw_projector_b}) which was derived for the projection operator.
This has the profound consequence that the semiclassical density of states which was originally derived in a wave packet approach~\cite{Xiao2005,Xiao2010} is not only determined by the projector condition, but it also follows from a general principle of gauge equivalence.

\subsection{Correction to the curvature}

By now we have established that the intraband terms of the projector $\rho$ are indeed fully determined by geometric and algebraic conditions, namely the Seiberg-Witten flow from Eq.~(\ref{eq:sw_projector_b}) and the projector condition $\rho \star \rho = \rho$.
Since the density matrix enters the Hall conductivity via the St\v{r}eda formula Eq.~(\ref{eq:streda}), it can be expected that similar flow terms arise also in the Berry curvature, possibly resembling the one in  Eq.~(\ref{eq:sw_curvature}). 
In order to determine if this is actually the case, we take the local Berry curvature in phase space which was previously defined as
\begin{equation}
\Omega_{\mu\nu} = \tr ~ 
      \rho \star  \Curv_{\mu\nu} ,
\end{equation}
and differentiate it with respect to $\hbar$. We again disregard the interband contributions from $\partial_\hbar \rho$ (which will vanish under the cyclic property of the trace) and take the semiclassical limit of $\hbar \to 0$ in order to obtain the first order correction $ \Delta \Omega_{\mu\nu}$, i.e., 
\begin{equation}
    \Omega_{\mu\nu}  - \left. \Omega_{\mu\nu} \right|_{\hbar \to 0} =  \Delta \Omega_{\mu\nu} + \mathcal{O}(\hbar^2) .
\end{equation}
The detailed derivation can be found in see appendix \ref{app:curvature} and reveals two distinct contributions: 
\begin{align}
    \Delta\Omega_{\mu\nu} & =  \hbar \left( -3 \tr ~ \varrho  (\partial_\hbar \Curv_{\mu\nu})^{\mathrm{SW}}_{\hbar\to0} + \mathcal{M}_{\mu\nu} \right),
    \label{eq:sw_hall}
\\
    \mathcal{M}_{\mu\nu} &= - \frac{1}{2}\Pi^{\kappa\lambda} \tr ~\varrho  \lbrace \partial^2_{\mu\kappa} \varrho, \partial^2_{\nu\lambda} \varrho \rbrace .
\label{eq:metric_term}
\end{align}
The first term in $\Delta \Omega_{\mu\nu}$ is indeed governed by the flow behavior as dictated by the Seiberg-Witten map of Eq.~(\ref{eq:sw_curvature}) which we have denoted explicitly by ``SW".
In contrast, it seems that the second term is not related to the Seiberg-Witten formalism in any way.
However, one has to keep in mind that the solution to the Seiberg-Witten equations is not unique and such a term might very well be in accordance with the general assumptions of the theory, although this is currently not known. 
Based on the symmetry of its second order derivatives, one can only tell that this term is intimately related to the quantum metric tensor of Eq.~(\ref{eq:quantum_metric_tensor}). 
In the following section, we further scrutinise this decomposition for the example of the two-dimensional Rashba model.

\figureII

\section{Noncollinear magnetism}
\label{sec:rashba}

\subsection{Rashba model}

In our recent work~\cite{Lux2020}, we have investigated the first order gradient correction to the Fermi sea contribution of the Hall conductivity in the two-dimensional Rashba model, whose Hamiltonian is given by
\begin{equation}
    H = \frac{\vec{p}^2}{2 m^\ast_\mathrm{e}} + \soi ( \vec{p} \times \bsigma)_z + \xc \hatn\cdot \bsigma.
    \label{eq:rashba_hamiltonian}
\end{equation}
Here, $m^\ast_\mathrm{e}$ is the effective electron mass, $\soi$ is the Rashba spin-orbit interaction and $\xc$ the strength of the exchange coupling to a noncollinear vector field $\hatn = \hatn(\vec{x})$.
Numerical spin dynamics simulations revealed that one-dimensional textures such as the conical spiral of the form
\begin{center}
    \includegraphics[width=0.95\linewidth]{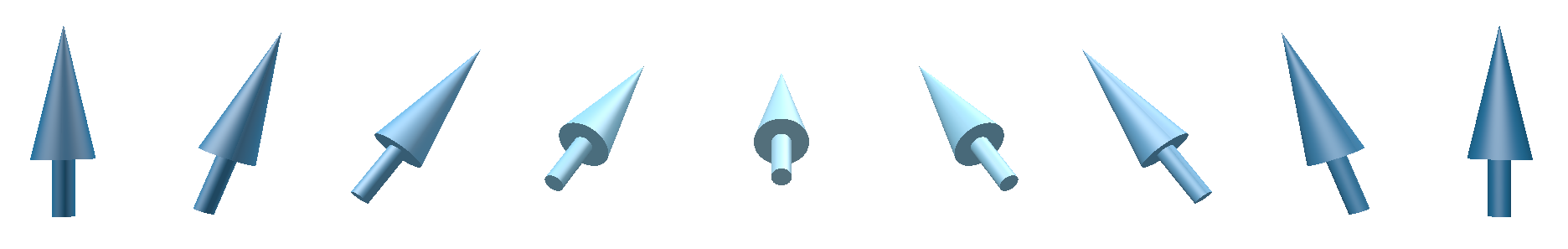} 
\end{center}
could provide a sizeable Hall signal in the first order term of the gradient expansion.
Another peculiar finding was the appearance of a coupling between the momentum space Berry curvature and the mixed space Berry curvature in the analytical expression for the gradient expansion whose origin remained uncertain at the time. 
The Seiberg-Witten flow of the curvature now provides an explanation of this term, which indeed emerges in Eq.~(\ref{eq:sw_curvature}).
We therefore want to revisit the Rashba model in the light of these findings.

\subsection{Band structure}

The band structure of the Rashba model is largely dominated by its topological character in momentum space.
Diagonalizing the Hamilontian Eq.~(\ref{eq:rashba_hamiltonian}) at a local position in real-space by treating the magnetization  direction $\hatn$ as a constant, one obtains the local band structure at this respective position which is visualized in Fig.~\ref{fig:bandstructure} a) for the model parameters $m^\ast_\mathrm{e} = m_\mathrm{e}$, $ \xc = \SI{0.5}{\electronvolt}$ and $\hbar\soi= \SI{1.95}{\electronvolt \angstrom}$.
If $n_z \neq 0$, one can assign the Chern number $\pm \sgn(n_z)/2$ to the respective bands which are nondegenerate for all momenta $\vec{k}$.
Since a sign change in $n_z$ also interchanges the Chern number between the bands, a topological phase transition occurs at $n_z = 0$ where the bands cross at a singular point $\hbar\vec{k}_c = (-n_y, n_x)^T \xc / \soi$. 
When being viewed in the context of a smoothly varying real-space magnetization texture, this singularity is also known as a mixed Weyl point~\cite{Hanke2017}.

This aspect is very important for the understanding the conical spiral phase which was alluded to above. 
In order to quantify our analysis, we parameterize this spin spiral as a rotation of the state $\hatn = \vec{e}_z$ by an angle $\psi = 2\pi x / \lambda$ with wavelength $\lambda$
around the axis $\boldsymbol{\omega}=\boldsymbol{\omega}(\theta, \phi)$, which is parameterized in spherical coordinates. 
The magnetization therefore traces a circular path when projected onto the unit sphere which is shown in Fig.~\ref{fig:bandstructure} b) for $\theta=\SI{0.6}{\radian}$ and $\phi=\SI{1.57}{\radian}$.

As a consequence, the system is driven close to its phase transition along the direction of the spiral which can be quantified by observing the band gap at the critical momentum $\vec{k}_c $, shown in Fig.~\ref{fig:bandstructure} c).
At $x=0.5\lambda$, the system is closest to its in-plane configuration $n_z=0$ which is never reached exactly. 
Since this type of almost-degeneracies typically presents the source of strong Berry curvature effects, one can expect a pronounced impact on the semiclassical density of states and the Hall coefficient in their vicinity.

\subsection{Energy resolved density of states}

Given the electron particle density $n_\mathrm{e}$ as function of the chemical potential $\mu$, we define the energy resolved density of states as
\begin{equation}
    \braket{\mathrm{dos}(\mu)} \equiv \frac{\partial n_\mathrm{e}}{\partial \mu}.
\end{equation}
The brackets $\Braket{\bullet}$ again indicate the average with respect to the real space configuration. 
We are interested in the local effects which are introduced by the Seiberg-Witten flow:
\begin{equation}
    \Delta \mathrm{dos}(\mu,\vec{x}) = \hbar ~ \int \frac{\dd^2 \vec{k}}{(2\pi)^2}\tr \left. \partial_\hbar \rho  \right|_{\hbar\to 0} .
    \label{eq:local_dos}
\end{equation}
Since the interband terms cancel out in this expression due to the cylic property of the trace, the Seiberg-Witten flow as dictated by Eq.~(\ref{eq:sw_projector_b}) is exact. Eq.~(\ref{eq:local_dos}) can be seen as the the first order contribution in a series expansion  which is given by 
\begin{equation}
    \braket{\mathrm{dos}(\mu)} - \braket{\mathrm{dos}(\mu)}_{\hbar\to0}
    =\braket{\Delta\mathrm{dos}(\mu)} + \mathcal{O}(\hbar^2) .
\end{equation}
This additional contribution is strongly tied to the noncollinear magnetism in the Rashba model which enters the Seiberg-Witten flow via its gradients in real-space.
Taking the model parameters and the conical spiral as before, the change in the local density of states $\Delta\mathrm{dos} $ at $\mu=0$ is shown in Fig.~\ref{fig:dos}.
In order to arrive at these results, all integrations are performed numerically (the details can be found in appendix \ref{app:computational_details}).
It displays a slight enhancement in the vicinity of the in-plane configuration which modulates an almost constant background value whose magnitude is comparable with the DOS of a free electron gas when the spiral wavelength is of order of $\lambda \approx \lambda_0 = \SI{1.95}{\angstrom}$ (and then decreases as $\lambda_0 / \lambda$ for larger wavelengths).
Since $\Delta\mathrm{dos}$ is fundamentally of first order in the gradients of the magnetization texture, it is sensitive to its chiral properties.
For example, simply inverting the wavevector from $2\pi/ \lambda$ to $-2 \pi/ \lambda$ will also change the sign of $\Delta\mathrm{dos}$ which therefore can serve as a proxy for the chirality of the spin distribution in experimental techniques such as scanning tunneling microscopy~\cite{Perini2019}.

\figureIII

\figureIV

\subsection{Anomalous Hall effect}

Based on Eq.~(\ref{eq:ahe_coeff}), the first order correction to the anomalous Hall coefficient can be calculated as
\begin{equation}
    \Delta \sigma_{xy}  =   \frac{e^2}{h}~  \frac{1}{2\pi } \int \dd^2\vec{k}~ \Delta \Omega_{k_x k_y}(\vec{k}),
\end{equation}
where $\Delta \Omega_{k_x k_y}$ is determined by Eqs.~(\ref{eq:sw_hall}).
Just as for the DOS, one can interpret this change as the average over the real space density $\Delta\sigma_{xy} \to \Braket{\Delta\sigma_{xy} (\vec{x})}$ and the local contributions can be investigated.
Since Eq.~(\ref{eq:ahe_coeff}) is composed of two distinct contributions, so is $\Delta\sigma_{xy}(\vec{x})$ and we refer to them as the Seiberg-Witten term (SW) and the metric term.
Both have first order terms with respect to gradients of the magnetization texture and are therefore sensitive to their chirality. 
For this reason, we have referred to the resulting effect as the \emph{chiral Hall effect} (CHE) in our recent work ~\cite{Lux2020}.
Fig.~\ref{fig:chiral_hall_effect} takes a look at the local evolution and the composition of the CHE along the direction of the conical spiral (for computational details we again refer to appendix \ref{app:computational_details}).
As the metric term vanishes, the response is solely determined by the SW term and reveals its strong sensitivity towards the change in the local band structure which is characterized by the influence of its adjacent mixed Weyl point. 
This is consistent with our previous findings in \cite{Lux2020} and now explains the observed response in the local conical phases as a consequence of the Seiberg-Witten map which in return is fueled by the proximity to a topological phase transition.

\section{Conclusion}

%-- general perspective
In this work, we have shown that the physics of noncollinear magnets can be reinterpreted in a fundamental way by embracing the noncommutative geometry of the underlying phase space.
For clean metals and insulators at low temperatures, this change of perspective reveals a relation between ordinary and noncommutative gauge transformations which influences the gradient expansion to all orders and whose mathematical structure was originally conceived based on low-energy approximations to string theory~\cite{Seiberg1999}. 
It is the great benefit of this new framework that it amends the gradient expansion technique with a geometric underpinning that is present from the very start.
This is to be seen in contrast to conventional Green's function techniques, where geometrical contributions have to be collected tediously~\cite{Lux2018}.
For the future, we see great promises for noncommutative geometry also beside this more traditional aspect of the gradient expansion. 

%-- rashba & SrRuO3
In application to the two-dimensional magnetic Rashba model, these new tools are able to explain our recent findings on a chirality dependent Hall response for the conical spiral phase~\cite{Lux2020} based on a geometrical interpretation.
The pronounced effect is induced by the Seiberg-Witten flow of the momentum space Berry curvature, which on its own is driven by an almost degenerate point in the band structure with a topological character. 
In the Rashba model, and in many realistic systems, this type of singularity is driven by the spin-orbit interaction~\cite{Hanke2017}.
Materials with a strong spin orbit coupling and noncollinear magnetism such as \ce{SrIrO3}/\ce{SrRuO3} bilayers~\cite{Matsuno2016, Ohuchi2018, Wysocki2020} hence form an interesting platform to study the inner workings of the Seiberg-Witten map. However, this formalism is not limited to the field of magnetism, but naturally extends to all cases where the electronic degrees of freedom depend on a smoothly varying order parameter.
This is the case for example in domain walls and topological structures in ferroelectric materials which received at lot of attention recently~\cite{Meier2015, Seidel2016, Seidel2019} as well as for smooth crystal deformations which can lead to a variety of geometrically induced effects~\cite{Yudin2013, Dong2018}.

%-- noncommutative promises
The ability of noncommutative geometry to analyze properties of strongly disordered systems is not unprecedented and has been used for example to study the quantum critical point of the transition between the trivial insulator and the quantum Hall insulator regime with a high degree of mathematical rigor~\cite{Prodan2013, Prodan2016}.
Applying the same level of rigor to the physics of noncollinear magnets therefore could elevate the field to a deeper understanding of the complex effects which emerge from spin textures that vary on mesoscopic length scales.

\section{Acknowledgements}

We  acknowledge  funding  under SPP 2137 ``Skyrmionics" (project  MO  1731/7-1)  of  Deutsche  Forschungsgemeinschaft (DFG) and also gratefully thank the J\"ulich Supercomputing Centre and RWTH Aachen University for providing computational resources under project jiff40.

%----------------------------------------------------
% literature
%----------------------------------------------------

\hbadness=99999 
%\bibliographystyle{apsrev4-1}
%\bibliography{literature}
%merlin.mbs apsrev4-1.bst 2010-07-25 4.21a (PWD, AO, DPC) hacked
%Control: key (0)
%Control: author (8) initials jnrlst
%Control: editor formatted (1) identically to author
%Control: production of article title (-1) disabled
%Control: page (0) single
%Control: year (1) truncated
%Control: production of eprint (0) enabled
%

%\clearpage

%\figureII

%\figureIII

%\figureIV

%\vfill
%\cleardoublepage

%----------------------------------------------------
\appendix
%----------------------------------------------------

\section{Derivation of the projector formula for metals and insulators}
\label{app:projector_formula}

In the clean limit, the electronic density matrix  is given by
\begin{equation}
    \rho \equiv - \frac{1}{\pi}\Im \int\limits_{-\infty}^\infty \dd \epsilon ~ n_\mathrm{F}(\epsilon)  ~\Gret =  \sum_{n } n_\mathrm{F}(\epsilon_n )f_{nn}(\vec{x}, \vec{p}) .
\end{equation}
At zero temperature limit, one either has $n_\mathrm{F}(\epsilon_n )=1$ or $n_\mathrm{F}(\epsilon_n )=0$ such that the density matrix fulfills the projector condition $\rho \star \rho = \rho$, which means that it describes a pure quantum state.
The density matrix of the holes is given accrodingly by the complement
$
    \bar{\rho} = \id - \rho 
$,
measuring the unoccupied subspace.
The off-diagonal elements of the conductivity tensor are then determined by~\cite{Lux2020}
\begin{align}
    \sigma^\mathrm{sea}_{xy} &= \hbar e^2\Re ~
    \fraktr_\mathrm{sea} \Braket{  \Gret \star v_x \star \Gret  \star v_y \star \Gret - ( x \! \leftrightarrow \! y) },
    \\
     \sigma^\mathrm{surf}_{xy} &= \hbar e^2  \Re ~
    \fraktr_\mathrm{surf} \Braket{
    v_x \star ( \Gret - \Gadv ) \star v_y \star \Gadv 
    },
    \notag \\
\end{align}
where the trace in the case of two space dimensions is given by
\begin{equation}
    \fraktr ( \bullet ) \equiv   \frac{1}{2\pi}  \int \dd\epsilon
    \int \frac{ \dd^2 \vec{p}}{(2 \pi \hbar )^{2}}~\tr ( \bullet ) .
\end{equation}
The two contributions $ \sigma^\mathrm{sea}_{xy}$ and $\sigma^\mathrm{surf}_{xy}$ add up to a transverse response which corresponds to an electric current in $x$-direction if an electric field is applied in the $y$-direction: $j_x = \sigma_{xy} E_y$. 
Assuming that the spectrum of the Hamiltonian is known, the expression can be simplified. 
For this purpose, it is useful to write the velocity operator in its commutator form, i.e.,
$
    v_i \equiv
    \partial_{p_i}H=  [x_i , H]_\star / (i \hbar)
$.
By an insertion of identity, one then finds
\begin{align}
    \Gret \star v_i  
        &=  \Gret \star v_i \star \id
    \nonumber \\ 
        &=  \sum_m \Gret \star v_i \star f_{mm}
    \nonumber \\ 
        &=  \frac{1}{i\hbar}\sum_{n,m\neq n}
            \frac{(\epsilon_m - \epsilon_n)}{\epsilon - \epsilon_n + i0^+} 
            ~ f_{nn} \star x_i \star f_{mm} ,
\end{align}
and similar for the advanced component of the Keldysh Green's function. In order to evaluate the Fermi surface contribution, one needs to make use of the distributional identity
\begin{align}
    g_n^\mathrm{A} (  g_m^\mathrm{R} - g_m^\mathrm{A} ) =- \frac{2 \pi i}{ \epsilon_m - \epsilon_n} \delta(\epsilon - \epsilon_m),
\end{align}
where $g_n^\mathrm{R/A} = (\epsilon - \epsilon_n \pm i0^+)^{-1}$ and where $\delta$ is the Dirac distribution. By defining the operator $
   c_{nm} \equiv   \int \dd^2 \vec{p}~\tr \braket{  f_{nn} \star x \star f_{mm} \star y}/ (2 \pi \hbar )^{2}$ and $\Delta_{mn} = \epsilon_m - \epsilon_n$, the Fermi surface term can be written as
\begin{align}
  &  \sigma^\mathrm{surf}_{xy} = \hbar e^2  \Re ~
    \fraktr_\mathrm{surf} \Braket{
   \Gadv \star v_x \star ( \Gret - \Gadv ) \star v_y
    }
    \notag
\\ &= \frac{e^2}{\hbar} \Re \sum_{nm} \frac{1}{2 \pi}  \int\limits_{-\infty}^\infty \dd\epsilon~ n_\mathrm{F}'(\epsilon)  g_n^\mathrm{A} (  g_m^\mathrm{R} - g_m^\mathrm{A} )\Delta_{mn}^2 c_{nm}     
\notag
\\ &= \frac{e^2}{\hbar} \sum_{n m} n_\mathrm{F}'(\epsilon_m) \Delta_{mn} \Im ~ c_{nm} .
\end{align}
In a similar way, one can proceed with the Fermi sea contribution, noting that $c_{nm} - (x \leftrightarrow y) = 2 i \Im ~c_{nm} $.
\begin{align}
    &  \sigma^\mathrm{sea}_{xy} = \hbar e^2\Re ~
    \fraktr_\mathrm{sea} \Braket{  \Gret \star v_x \star \Gret  \star v_y \star \Gret - ( x \! \leftrightarrow \! y) }
\notag \\
& = -\frac{e^2}{\hbar}  \sum_{nm} \frac{1}{ \pi}\Im  \int\limits_{-\infty}^\infty \dd\epsilon~ n_\mathrm{F}(\epsilon)  (g_n^\mathrm{R})^2   g_m^\mathrm{R} \Delta_{mn}^2 \Im ~c_{nm}     
\end{align}
The energy integral evaluates to two different contributions:
\begin{align}
     \frac{1}{\pi} \Im \int \dd \epsilon ~ n_\mathrm{F}(\epsilon)\Delta_{mn}^2 g_n^2 g_m 
    =& (n_\mathrm{F}(\epsilon_n) - n_\mathrm{F}(\epsilon_m) ) + \frac{n_\mathrm{F}'(\epsilon_n)}{\Delta_{mn}}  .
\end{align}
The second term in this identity will therefore lead to a contribution
\begin{align}
 \Delta \sigma^\mathrm{sea}_{xy}  &= -\frac{e^2}{\hbar}  \sum_{nm}
  n_\mathrm{F}'(\epsilon_n) \Delta_{mn} \Im ~c_{nm}
   \notag \\
  & = -\frac{e^2}{\hbar}  \sum_{nm} n_\mathrm{F}'(\epsilon_m) \Delta_{nm} \Im ~c_{mn} \notag \\
     & = -\frac{e^2}{\hbar}  \sum_{nm} n_\mathrm{F}'(\epsilon_m) \Delta_{mn} \Im ~c_{nm} = -\sigma^\mathrm{surf}_{xy},
\end{align}
where we have first interchanged the dummy indices and then used the antisymmetry of $\Delta_{nm}$ and $c_{nm}$. 
As this contribution is completely cancelled by the Fermi surface term, only the following (intrinsic) contribution remains:
\begin{align}
    &  \sigma_{xy} = -\frac{e^2}{\hbar}  \sum_{nm} (n_\mathrm{F}(\epsilon_n) - n_\mathrm{F}(\epsilon_m) ) \Im ~c_{nm}    
    \notag \\
    & = 
    -\frac{e^2}{\hbar}\Im  \int \frac{\dd^2 \vec{p}}{(2 \pi\hbar  )^{2}}~\sum_{nm} (n_\mathrm{F}(\epsilon_n) - n_\mathrm{F}(\epsilon_m) )
    \notag \\ & \hspace{2cm} \times
    ~\tr \braket{  f_{nn} \star x \star f_{mm} \star y} 
    \notag \\
    & = 
    -\frac{e^2}{\hbar}\Im  \int \frac{\dd^2 \vec{p}}{(2 \pi\hbar  )^{2}}~
    ~\tr \braket{  \rho \star x  \star \bar{\rho} \star y}  - (x\! \leftrightarrow \! y).
\end{align}
At finite temperatures, the density matrix describes a mixed state for a metallic system. A pure state is recovered only in the zero temperature limit, and $\rho \star \rho = \rho$ is fulfilled, while $\rho \star \bar{\rho} = 0$. Therefore,
\begin{align}
    \rho \star  x \star \bar{\rho} \star y \star \rho
&= \rho \star [x, \bar{\rho}]_\star  \star y \star \rho 
\notag \\
&= -\rho \star [x, \bar{\rho}]_\star  \star [y, \bar{\rho}]_\star \star \rho 
\notag \\
& = \hbar^2  \rho \star \partial_{p_x}\rho \star \partial_{p_y} \rho \star \rho.
 \end{align}
Inserting relation gives the desired result:
\begin{align}
      \sigma_{xy} 
    &=
    \frac{e^2}{h}   \frac{1}{2\pi}\int \dd^2 \vec{k}~
    ~i \tr \braket{\rho \star [ \partial_{k_x}\rho , \partial_{k_y} \rho ]_\star }.
\end{align}
% If the projector condition is relaxed, a second contribution arises which has the form
% \begin{align}
%       \delta\sigma_{xy} 
%     &=
%     -\frac{e^2}{\hbar}   \int~
%     \frac{ \dd^2 \vec{k}}{(2\pi)^2}
%     ~ \braket{ \vec{r} \times  \tr~ \rho \star \nabla_\vec{k}\rho \star \rho  }_z.
% \end{align}

%----------------------------------------------------
\section{Flow equation of the projection operator}
\label{app:projection_flow}
%----------------------------------------------------

In order to determine the gradient corrections to the curvature, one needs to determine the corrections to the projection operator. By construction, the projection operator fulfills
$
    \rho \star \rho = \rho .
$
In the limit of $\hbar\to 0$, this equation reduces to $\varrho \varrho = \varrho$. These requirements are quite strict. Differentiating with respect to $\hbar$ yields
\begin{equation}
    \partial_\hbar \rho = \lbrace \partial_\hbar \rho , \rho \rbrace_\star + \frac{i}{2} \Pi^{\mu\nu}  \partial_\mu \rho \star \partial_\nu \rho .
    \label{eq:sw_projector_app}
\end{equation}
Since $\rho\star \partial_\mu \rho \star \rho = 0$, the intraband terms are readily determined:
\begin{eqnarray}
       \rho \star \partial_\hbar \rho \star \rho 
        = &   -  \frac{i}{2} \Pi^{\kappa\lambda} \rho\star \partial_\kappa \rho \star \partial_\lambda \rho  
       \\
       \rho^\perp \star \partial_\hbar \rho \star \rho^\perp 
       = & +  \frac{i}{2} \Pi^{\kappa\lambda}\rho^\perp \star  \partial_\kappa \rho  \star \partial_\lambda \rho  .
\end{eqnarray}
Here we defined $\rho^\perp = \id - \rho$. 
The interband terms $ \rho \star \partial_\hbar \rho \star \rho^\perp $ and $ \rho^\perp \star \partial_\hbar \rho \star \rho $ are not directly determined by Eq.~\ref{eq:sw_projector_app}. 
Discarding these terms then yields
\begin{align}
    \partial_\hbar \rho     =   \frac{i}{2} \Pi^{\kappa\lambda} [\partial_\kappa \rho, \rho]_\star \star  \partial_\lambda \rho = - \frac{1}{2} \Pi^{\kappa\lambda} \Conn_\kappa \star \partial_\lambda \rho.
\end{align}
This result can be slightly rewritten using the identity :
\begin{align}
& - \frac{1}{4} \Pi^{\kappa \lambda}  \partial_\lambda \rho  \star \Conn_\kappa 
\notag \\
& =  \frac{1}{4} \Pi^{\kappa \lambda}  \partial_\kappa \rho \star  \Conn_\lambda
\notag \\ 
& = 
-  \frac{1}{4} \Pi^{\kappa \lambda} [ \rho, \Conn_\kappa ]_\star \star [\partial_\lambda \rho, \rho]_\star 
\notag \\ 
& = 
-  \frac{1}{4} \Pi^{\kappa \lambda} ( \rho + \rho^\perp)  \star \Conn_\kappa \star  \partial_\lambda \rho\star( \rho + \rho^\perp)
\notag \\ 
& = 
-  \frac{1}{4} \Pi^{\kappa \lambda}  \Conn_\kappa \star \partial_\lambda \rho,
\end{align}
which is valid in the parallel transport gauge, where $\Conn$ has no interband terms.
Inserting this result, and using the fact that $\nabla_\mu \rho =0$, one obtains the desired result
 \begin{align}
    \partial_\hbar \rho     =   - \frac{1}{4} \Pi^{\kappa\lambda} \lbrace \Conn_\kappa, \partial_\lambda \rho  + \nabla_\lambda \rho \rbrace_\star .
\end{align}

%----------------------------------------------------
\section{Correction of the Green's function}
\label{app:green_projector_formula}
%----------------------------------------------------

The correction to the Green's function is determined from the Dyson equation
\begin{equation}
    (\Gk_0)^{-1} \star \Gk   = \id 
\end{equation}
and is therefore generated by the differential flow equation
\begin{equation}
   \partial_\hbar \Gk = \frac{i}{2} \Pi^{\mu\nu} ~\Gk \star \partial_\mu  \Gk^{-1\star} \star \Gk \star \partial_\nu  \Gk^{-1\star} \star \Gk . 
\end{equation}
At zero temperature and in the clean limit, the projection operator of an insulator can be obtained from the Green's function using the residue formula
\begin{equation}
    \rho = \frac{1}{2\pi i } \oint\limits_\gamma ~\dd \epsilon~ \Gk (\epsilon),
\end{equation}
where the contour $\gamma$ encloses the real energy axis up to the Fermi energy in an anticlockwise direction.
We use the correspondence $ \partial_\mu = -i ~ \ad_\star ~\bar{z}_\mu$   to rewrite the flow equation in terms of derivations:     
\begin{equation}
   \partial_\hbar \Gk = -\frac{i}{2} \Pi^{\mu\nu} ~\Gk \star \ad_ {\bar{z}_\mu}^\star \Gk^{-1\star} \star \Gk \star \ad^\star_{\bar{z}_\nu} \Gk^{-1\star} \star \Gk . 
\end{equation}
This equation can be written in terms of matrix elements
\begin{align}
  &  f_{nn} \star \partial_\hbar \Gk \star f_{mm} 
 \notag \\
 & = 
    -\frac{i}{2} \Pi^{\mu\nu} ~(g_n  (\ad_ {\bar{z}_\mu}^\star \Gk^{-1\star})_{np} \star g_p  (\ad^\star_{\bar{z}_\nu} \Gk^{-1\star})_{pm}  g_m ) f_{nm} 
\notag\\ & \equiv (\partial_\hbar \Gk )_{nm} f_{nm},
\notag \\
\end{align}
where a summation over repeated indices is implied.
Here we have
\begin{align}
    (\ad_ {\bar{z}_\mu}^\star \Gk^{-1\star})_{np} 
    &=( [ \bar{z}_\mu, \Gk^{-1\star}]_\star )_{np} 
    \notag \\
    &= ( [H,  \bar{z}_\mu]_\star )_{np} =  (\epsilon_n - \epsilon_p)
    ( \bar{z}_\mu)_{np} 
    \notag \\
    &= \Delta_{np}  ( \bar{z}_\mu)_{np} .
\end{align}
From which follows that
\begin{align}
    (\partial_\hbar \rho)_{nm} & =  -\frac{i}{2} \Pi^{\mu\nu}  \sum_{p} \Delta_{np} \Delta_{pm}  ( \bar{z}_\mu)_{np} ( \bar{z}_\nu)_{pm} \mathcal{I}_{npm}.
\end{align}
Here, the contour integrations are summarized in the quantity
\begin{equation}
   \mathcal{I}_{npm} \equiv   \frac{1}{2\pi i }  \oint\limits_\gamma ~\dd \epsilon ~ g_n g_p g_m .
\end{equation}
For $n\neq m$, Cauchy's residue theorem gives 
\begin{align}
 \mathcal{I}_{npm}
  &= \theta( \mu - \epsilon_n) \mathrm{Res}( g_p g_m; \epsilon_n) + \mathrm{cyclic}
  \notag  \\
  &=  \frac{\theta( \mu - \epsilon_n)}{\Delta_{np} \Delta_{nm}} + \frac{\theta( \mu - \epsilon_p)}{\Delta_{pn} \Delta_{pm}}
  + \frac{\theta( \mu - \epsilon_m)}{\Delta_{mn} \Delta_{mp}}
  \notag  \\
  &=  \frac{1}{\Delta_{np} \Delta_{pm}} \left( 
  - \theta( \mu - \epsilon_p) +  \vartheta_{npm}
  \right),
\end{align}
where
\begin{equation}
    \vartheta_{npm} \equiv \frac{1}{\Delta_{nm}} \left( \Delta_{pm}\theta( \mu - \epsilon_n) - \Delta_{pn}\theta( \mu - \epsilon_m)\right).
\end{equation}
In the degenerate limit of $n\to m$, $\vartheta_{npm}$ becomes a Fermi surface contribution and can be neglected for insulators. 
Observe that
\begin{align}
    & (\partial_\mu \rho)_{nm} f_{nm} =  f_{nn} \star \partial_\mu \rho \star f_{mm}
    \notag \\
    & = - i \sum_{ \epsilon_p < \mu}  f_{nn} \star \ad_{\bar{z}_\mu}^\ast f_{pp} \star f_{mm}
    \notag\\
    & = + i ( \bar{z}_\mu )_{nm} ( \theta( \mu - \epsilon_n) -\theta( \mu - \epsilon_m) ) ~ f_{nm}
\end{align}
If $\epsilon_n < \mu$ and $\epsilon_m > \mu$ this implies
\begin{equation}
    ( \bar{z}_\mu )_{nm} = -i  (\partial_\mu \rho)_{nm} .
\end{equation}
If $\epsilon_m < \mu$ and $\epsilon_n > \mu$ this implies
\begin{equation}
    ( \bar{z}_\mu )_{nm} = +i  (\partial_\mu \rho)_{nm} .
\end{equation}
Assume that $\epsilon_n > \mu$ and $\epsilon_m > \mu$. Then
\begin{align}
    (\partial_\hbar \rho)_{nm} & =  + \frac{i}{2} \Pi^{\mu\nu}  \sum_{\epsilon_p < \mu}  ( \bar{z}_\mu)_{np} ( \bar{z}_\nu)_{pm} 
    \notag \\
    & =  + \frac{i}{2} \Pi^{\mu\nu}  \sum_{\epsilon_p < \mu}  ( \partial_\mu \rho )_{np} ( \partial_\nu \rho)_{pm} 
\end{align}
And therefore
\begin{equation}
    \rho^\perp \star\partial_\hbar \rho \star  \rho^\perp =   \frac{i}{2} \Pi^{\mu\nu}  \partial_\mu \rho \star \rho \star \partial_\nu \rho,
\end{equation}
in full agreement with the previously obtained result. If $\epsilon_n < \mu$ and $\epsilon_m < \mu$ one has
\begin{align}
     & - \theta( \mu - \epsilon_p) +  \frac{1}{\Delta_{nm}} \left( \Delta_{pm}\theta( \mu - \epsilon_n) - \Delta_{pn}\theta( \mu - \epsilon_m)\right)  
     \notag \\&
     = 1 - \theta( \mu - \epsilon_p) ,
\end{align}
therefore $\epsilon_p > \mu$ in the summation and one obtains
\begin{equation}
    \rho \star\partial_\hbar \rho \star  \rho =   -\frac{i}{2} \Pi^{\mu\nu}  \partial_\mu \rho \star \rho^\perp \star \partial_\nu \rho,
\end{equation}
which is also in agreement with the previous derivations.  Now to the offdiagonal terms which could not be derived from the projection constraint. Assume that $\epsilon_n > \mu$ and $\epsilon_m < \mu$. Then
\begin{align}
     & - \theta( \mu - \epsilon_p) +  \frac{1}{\Delta_{nm}} \left( \Delta_{pm}\theta( \mu - \epsilon_n) - \Delta_{pn}\theta( \mu - \epsilon_m)\right)
      \\ & 
      =
      - \theta( \mu - \epsilon_p) - \frac{\Delta_{pn}}{\Delta_{nm}}
\end{align}
Inserting this result gives
\begin{align}
    (\partial_\hbar \rho)_{nm} & =  \frac{i}{2} \Pi^{\mu\nu}  \sum_{p}   ( \bar{z}_\mu)_{np} ( \bar{z}_\nu)_{pm} (\theta( \mu - \epsilon_p) + \frac{\Delta_{pn}}{\Delta_{nm}} ) 
    \notag \\ 
     & =  \frac{i}{2} \Pi^{\mu\nu}  \sum_{\epsilon_p > \epsilon_F}   ( \bar{z}_\mu)_{np} ( \bar{z}_\nu)_{pm} \frac{\Delta_{pn}}{\Delta_{nm}}
     \notag \\ &
     +
      \frac{i}{2} \Pi^{\mu\nu}  \sum_{\epsilon_p < \epsilon_F}   ( \bar{z}_\mu)_{np} ( \bar{z}_\nu)_{pm} \frac{\Delta_{pm}}{\Delta_{nm}} .
%     \notag   \\
  %   & =  \frac{i}{2} \Pi^{\mu\nu}  \frac{1}{\Delta_{nm}} (\rho^\perp \star \lbrace \partial_\mu \rho, \partial_\nu H \rbrace_\star \star \rho )^{nm}
\end{align}
There seems to be no straightforward way to express this as an operator product. 
One sees however that in the adiabatic limit of a large gap $\Delta E$, these interband terms decay as $\mathcal{O}(1/\Delta E)$ since $ | \Delta_{nm} | \geq \Delta E$.
Further, the interband terms cannot contribute to the correction of the Hall conductivity.

\section{Correction to the curvature}
\label{app:curvature}

The deformed Berry curvature is defined as
\begin{equation}
\Omega_{\mu\nu} \equiv \tr ~ 
      \rho \star  \Curv_{\mu\nu} ,
\end{equation}
and we need to differentiate it with respect to $\hbar$. 
To simplify this analysis, we again disregard the interband contributions from $\partial_\hbar \rho$ (which will not contribute anyway) and take the semiclassical limit of $\hbar \to 0$ in order to obtain the first order correction $ \Delta \Omega_{\mu\nu}$. 
Several terms appear, while the simplest one of those is given by
\begin{equation}
    \Delta\Omega_{\mu\nu}^{(1)} \equiv  \hbar \tr \left. ( \partial_\hbar \rho  \star \Curv_{\mu\nu}) \right|_{\hbar\to 0}.
\end{equation}
To simplify the notation, we introduce the limits
\begin{align}
    \conn_\mu & = \lim\limits_{\hbar\to 0} \Conn_\mu , \\
    \curv_\mu & = \lim\limits_{\hbar\to 0} \Curv_\mu .
\end{align}
By inserting the flow equation of the projector, one finds
\begin{align}
     \Delta\Omega_{\mu\nu}^{(1)}/ \hbar &
 =   \frac{i}{2} \Pi^{\kappa\lambda}  \tr~ [\partial_\kappa \varrho, \varrho]  \partial_\lambda \varrho   \curv_{\mu\nu} 
\notag \\ 
& =    -\frac{1}{2} \Pi^{\kappa\lambda}  \tr~ \curv_{\kappa\lambda}   \varrho  \curv_{\mu\nu} 
+ \frac{1}{2} \Pi^{\kappa\lambda}  \tr~ \conn_\kappa  \varrho  \partial_\lambda \curv_{\mu\nu} 
\notag \\ 
& =   -\frac{1}{4} \Pi^{\kappa\lambda}  \tr~ \varrho\lbrace  \curv_{\kappa\lambda} ,  \curv_{\mu\nu} \rbrace
+ \frac{1}{2} \Pi^{\kappa\lambda}  \tr~ \conn_\kappa  \varrho  \partial_\lambda \curv_{\mu\nu} 
\notag \\
& =  \frac{1}{4}\Pi^{\kappa\lambda}   \tr~\varrho\left( \lbrace \conn_\kappa, \partial_\lambda \curv_{\mu\nu}\rbrace  - 2\lbrace  \curv_{\kappa\mu} ,  \curv_{\lambda\nu} \rbrace
\right)  .
\end{align}
These results can be summarized to
\begin{align}
\Delta\Omega_{\mu\nu}^{(1)}/\hbar = & 
   -\tr ~ \varrho  (\partial_\hbar \Curv_{\mu\nu} )^\mathrm{SW}_{\hbar\to0}
   - \frac{1}{4}\Pi^{\kappa\lambda}   \tr~\varrho\lbrace \conn_\kappa, \nabla_\lambda \curv_{\mu\nu}\rbrace  ,
\end{align}
where the superscript ``SW" indicates that the $\hbar$-flow is generated according to the Seiberg-Witten map. 
In the parallel transport gauge, the second term is zero for the covariant derivative $\nabla_\mu \bullet= \varrho  (\partial_\mu\bullet)  \varrho$. 
A second category of terms originates in the $\hbar$-derivative of the $\star$-product itself:
\begin{equation}
    \Delta\Omega_{\mu\nu}^{(2)} \equiv \hbar \Tr ~  \varrho (\partial_\hbar \star)_{\hbar \to 0} \curv_{\mu\nu}.
\end{equation}
This term vanishes due to the antisymmetry of the symplectic structure in combination with the cyclic property of the trace under the phase space integral.
The last category of terms stems directly from the curvature:
\begin{equation}
    \Delta\Omega_{\mu\nu}^{(3)} \equiv \hbar\tr ~  \varrho (\partial_\hbar \Curv_{\mu\nu})_{\hbar\to 0}.
\end{equation}
Some of the contributions can be related to $\Delta\Omega_{\mu\nu}^{(3)}$ via partial integration:
\begin{align}
    & i \tr \varrho  [ \partial_\mu \partial_\hbar \varrho, \partial_\nu \varrho] 
    +
    i \tr \varrho  [ \partial_\mu  \varrho, \partial_\nu \partial_\hbar\varrho] 
    \notag \\
     & =   - i \tr \partial_\mu \varrho  [ \partial_\hbar \varrho, \partial_\nu\varrho] 
    - i \tr \varrho  [ \partial_\hbar \varrho, \partial_\nu\partial_\mu \varrho]
    \notag \\
    & \phantom{=} - i \tr \partial_\nu\varrho  [ \partial_\mu  \varrho, \partial_\hbar\varrho]  - i \tr \varrho  [ \partial_\nu\partial_\mu  \varrho, \partial_\hbar\varrho] 
     \notag \\
     & =   - i \tr \partial_\mu \varrho  [ \partial_\hbar \varrho, \partial_\nu\varrho] 
    - i \tr \partial_\nu\varrho  [ \partial_\mu  \varrho, \partial_\hbar\varrho]  
    \notag \\
     & = + 2 i \tr \partial_\hbar \varrho [\partial_\mu \varrho  , \partial_\nu\varrho] 
    \notag \\
     & = 2  \Delta\Omega_{\mu\nu}^{(1)}/\hbar .
\end{align}
The partial integration relies on the fact that $\partial_\hbar \varrho$ is assumed to have no interband contributions. 
The last derivative term is then given by
\begin{equation}
    \Delta\Omega_{\mu\nu}^{(3)} = 2  \Delta\Omega_{\mu\nu}^{(1)} - \frac{\hbar}{2}\Pi^{\kappa\lambda} \tr ~\varrho  \lbrace \partial^2_{\mu\kappa} \varrho, \partial^2_{\nu\lambda} \varrho \rbrace . 
\end{equation}
The sum $\Delta\Omega_{\mu\nu}^{(1)}+\Delta\Omega_{\mu\nu}^{(2)}+\Delta\Omega_{\mu\nu}^{(3)}$ combines to the total derivative 
\begin{align}
    \Delta\Omega_{\mu\nu} & = \hbar \left( -3 \tr ~ \varrho  (\partial_\hbar \Curv_{\mu\nu})^{\mathrm{SW}}_{\hbar\to0} + \Delta_{\mu\nu} \right),
\end{align}
with the additional term
\begin{align}
    \Delta_{\mu\nu} =& - \frac{1}{2}\Pi^{\kappa\lambda} \tr ~\varrho  \lbrace \partial^2_{\mu\kappa} \varrho, \partial^2_{\nu\lambda} \varrho \rbrace .
\end{align}

\section{Computational details}
\label{app:computational_details}

All momentum space integrations have been performed on a uniform mesh of $3000\times3000$ $k$-points. 
Since the model is continuous, a cutoff is introduced with a magnitude of approximately $k_\mathrm{cut} \approx 2.5~k_\mathrm{F}$, where $k_\mathrm{F}$ is the largest Fermi momentum. 
States with higher momenta are then sure to be unoccupied and do not contribute to the integration.
All derivatives are evaluated with finite differences.

%\vfill
%\clearpage

\end{document}